\documentclass[draftclsnofoot, onecolumn, 12pt]{IEEEtran}
\usepackage{amsmath,graphicx}
\usepackage{algorithm}
\usepackage{algpseudocode}
\usepackage{amssymb}
\usepackage{subfigure}
\usepackage{footmisc}
\usepackage{lipsum}
\usepackage{mathtools}
\usepackage{cuted}
\usepackage{esint}
\usepackage{multicol}
\usepackage{multirow}
\usepackage{cite}
\usepackage{xcolor}
\usepackage{accents}
\usepackage{etoolbox}

\makeatletter
\@for\next:={int,iint,iiint,iiiint,dotsint,oint,oiint,sqint,sqiint,
  ointctrclockwise,ointclockwise,varointclockwise,varointctrclockwise,
  fint,varoiint,landupint,landdownint}\do{%
    \expandafter\edef\csname\next\endcsname{
      \noexpand\DOTSI
      \expandafter\noexpand\csname\next op\endcsname
      \noexpand\ilimits@
    }%
  }
\makeatother

\begin{document}

\title{Performance Analysis of Active Large Intelligent Surfaces (LISs): Uplink Spectral Efficiency and Pilot Training}

\author{Minchae~Jung,~\IEEEmembership{Member,~IEEE}, Walid~Saad,~\IEEEmembership{Fellow,~IEEE},
and Gyuyeol~Kong
\thanks{

A preliminary version of this work was submitted to IEEE GLOBECOM 2019 \cite{ref.Jung2019PCcon}.

{M. Jung is with the Department of Electronic Engineering, Soonchunhyang University, Asan, Chungcheongnam-do, Rep. of Korea (e-mail: hosaly@sch.ac.kr).}

{W. Saad is with Wireless@VT, Department of Electrical and Computer Engineering, Virginia Tech, Blacksburg, VA 24061 USA (e-mail: walids@vt.edu).}

{G. Kong is with Department of Signal Processing and Acoustics, Aalto University, Finland (e-mail: gyuyeol.kong@aalto.fi).}

This research was supported by Basic Science Research Program through the National Research Foundation of Korea (NRF) funded by the Ministry of Education (NRF-2016R1A6A3A11936259)
and by the U.S. National Science Foundation under Grants CNS-1836802, CNS-1617896, and OAC-1638283.\vspace{-1.7cm}
}
}

\markboth{}%
{Shell \MakeLowercase{\textit{et al.}}: Bare Demo of IEEEtran.cls for Journals}

\maketitle \vspace{-1.5cm}
\begin{abstract} \vspace{-0.2cm}
Large intelligent surfaces (LISs) constitute a new and promising wireless communication paradigm that relies on the integration of a massive number of antenna elements over the entire surfaces of man-made structures.
The LIS concept provides many advantages, such as the capability to provide reliable and space-intensive communications by effectively establishing line-of-sight (LOS) channels.
In this paper, the system spectral efficiency (SSE) of an active LIS system is asymptotically analyzed under a practical LIS environment with a well-defined uplink frame structure.
In order to verify the impact on the SSE of pilot contamination, the SSE of a multi-LIS system is asymptotically studied and a theoretical bound on its performance is derived.
Given this performance bound, an optimal pilot training length for multi-LIS systems subjected to pilot contamination is characterized and, subsequently, 
the number of devices that need to be serviced by the LIS in order to maximize the performance is derived.
Simulation results show that the derived analyses are in close agreement with the exact mutual information in presence of a large number of antennas,
and the achievable SSE is limited by the effect of pilot contamination and intra/inter-LIS interference through the LOS path, even if the LIS is equipped with an infinite number of antennas.
Additionally, the SSE obtained with the proposed pilot training length and number of scheduled devices is shown to reach the one obtained via a brute-force search for the optimal solution.
\end{abstract}

\vspace{-0.1cm}
\begin{IEEEkeywords}\vspace{-0.3cm}
Large intelligent surface (LIS), large system analysis, performance analysis, pilot contamination, system spectral efficiency, 

\end{IEEEkeywords}

\IEEEpeerreviewmaketitle
\section{Introduction}
\IEEEPARstart{T}{he} notion of a large intelligent surface (LIS) that relies on equipping man-made structures, such as buildings, with massive antenna arrays is rapidly emerging as a key enabler of tomorrow's Internet of Things (IoT) and sixth generation (6G) applications \cite{ref.Basar2019indexmodulation,ref.Saad20196G, ref.Ericsson2011IoT, ref.Dawy2017MTC, ref.Tae2016learning,ref.Mozaffari2019beyond,ref.Mozaffari2017mobile, ref.Zeng2018joint,ref.Jung2018lisul, ref.Hu2018data, ref.Hu2018assignment, ref.Jung2019reliability, ref.Hu2018positioning,ref.Hu2018hardware ,ref.LISnew,ref.Han2018assisted, ref.Huang2018energy,ref.Wu2018beamforming, ref.Wu2019bfoptimization,ref.editor,ref.PSS00,ref.WPT00,ref.WPT01,ref.RR1}.
An LIS system can potentially provide pervasive and reliable wireless connectivity by exploiting the fact that pervasive city structures, such as buildings, roads, and walls, will become electromagnetically active in the near future.
If properly operated and deployed, the entire environment is expected to be active in wireless transmission providing near-field communications.
In contrast, conventional massive multiple-input multiple-output (MIMO) systems is essentially regarded as far-field communications generating non-line-of-sight (NLOS) channels with a high probability.
Indeed, the wireless channels of an LIS can become nearly line-of-sight (LOS) channels, resulting in several advantages compared to conventional massive MIMO system.
First, noise and inter-user interference through a NLOS path become negligible as the number of antenna arrays on LIS increases \cite{ref.Jung2018lisul}.
Also, the inter-user interference through a LOS path is negligible providing an interference-free environment, when the distances between adjacent devices are larger than half the wavelength \cite{{ref.Hu2018data},{ref.Hu2018assignment}}.
Moreover, an LIS offers more reliable and space-intensive communications compared to conventional massive MIMO systems as clearly explained in \cite{ref.Jung2018lisul} and \cite{ref.Jung2019reliability}.
Note that those advantages and differences between massive MIMO and LIS have been established in the prior literature and the interest reader is referred to \cite{ref.Jung2018lisul, ref.Hu2018data, ref.Hu2018assignment, ref.Jung2019reliability, ref.Hu2018positioning,ref.Hu2018hardware, ref.Han2018assisted, ref.Huang2018energy,ref.Wu2018beamforming, ref.Wu2019bfoptimization,ref.editor,ref.PSS00,ref.WPT00,ref.WPT01,ref.LISnew}.

\subsection{Prior Art}
Large antenna arrays can be either an LIS with active elements or an intelligent reflecting surface (IRS) with passive elements.
The traditional concept of an LIS is an extension of massive MIMO from a discrete array to a continuous surface
and, thus, it is practically a compact integration of massive number of tiny active antennas \cite{ref.Jung2018lisul, ref.Hu2018data, ref.Hu2018assignment, ref.Jung2019reliability, ref.Hu2018positioning,ref.Hu2018hardware,ref.LISnew}.
As a concept beyond massive MIMO, an LIS exploits the whole contiguous surface for transmitting and receiving information, 
which makes it possible to work as either a transmitter or receiver.
\begin{figure}[]
\centering\vspace{-3cm}
\includegraphics[width=1.0\columnwidth] {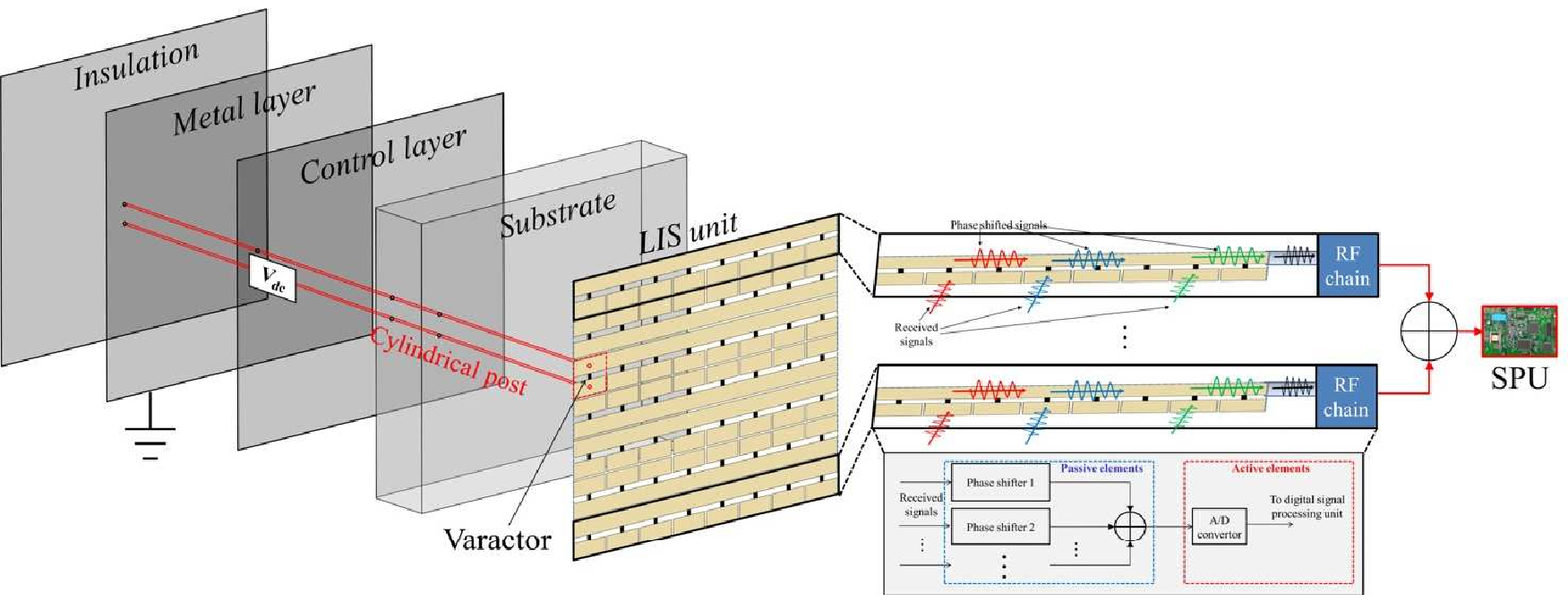}\vspace{-2.8cm}
\caption{Illustrative architecture of considered active LIS.}\vspace{-0.8cm}
\label{fig.01}
\end{figure}
In contrast, an IRS is regarded as a passive reflecting surface which manipulates the phase of an incident radio frequency (RF) signal from a base station (BS) and reflects it to the desired user using passive elements \cite{ref.Han2018assisted, ref.Huang2018energy,ref.Wu2018beamforming,ref.Wu2019bfoptimization,ref.editor,ref.PSS00,ref.WPT00,ref.WPT01}. 
Both of these cases of intelligent surface-based communication have received significant attention in the literature \cite{ref.Jung2018lisul, ref.Hu2018data, ref.Hu2018assignment, ref.Jung2019reliability, ref.Hu2018positioning,ref.Hu2018hardware,ref.LISnew, ref.Han2018assisted, ref.Huang2018energy,ref.Wu2018beamforming,ref.Wu2019bfoptimization
,ref.editor,ref.WPT00,ref.WPT01,ref.PSS00}.
In particular, the works in \cite{ref.Jung2018lisul} and \cite{ref.Hu2018data} provided an analysis of the uplink data rate to evaluate LIS performance considering channel estimation errors, and studied the space normalized capacity achieved by an optimal receiver and a matched filter (MF), respectively.
Also, in \cite{ref.Hu2018assignment} and \cite{ref.Jung2019reliability}, the authors proposed an optimal user assignment scheme to select the best LIS unit and analyzed the reliability of an LIS system in terms of the outage probability, respectively. 
Moreover, the works in \cite{ref.Hu2018positioning} and \cite{ref.Hu2018hardware} derived, respectively, the Fisher-information and Cramer-Rao lower bound for user positions exploiting the LIS uplink signal and the uplink capacity considering hardware impairments. 
In \cite{ref.LISnew}, the authors analyzed LIS performance in terms of the array gain, the spatial resolution, and the capability of interference suppression.
Meanwhile, the authors in \cite{ref.Han2018assisted} and \cite{ref.Huang2018energy} designed a beamformer and a phase shifter that maximize the ergodic rate and the energy efficiency, respectively.
Also, the works in \cite{ref.Wu2018beamforming} and \cite{ref.Wu2019bfoptimization} designed joint active and passive beamformers that minimize the transmit power at the BS, under, respectively, continuous and discrete phase shifts in an IRS-assisted system.
In \cite{ref.RR2}, the authors proposed a practical channel estimation framework in IRS-assisted system
and a prototype for a high-gain and low-cost IRS having 256 elements is developed in \cite{ref.editor}.
Also, IRSs have been used for some new applications such as physical layer security and simultaneous wireless information and power transfer (SWIPT) (e.g., see \cite{ref.PSS00,ref.WPT00, ref.WPT01}).
Furthermore, IRSs have been investigated with promising machine learning
techniques, such as a deep reinforcement learning \cite{ref.RR3} and unsupervised learning \cite{ref.RR4}.
More recently, the concept of a dynamic metasurface antenna (DMA) has emerged as a promising technology that can exploit the advantages of IRS based on the passive elements,
while still delivering active LIS features \cite{ref.DMA1, ref.DMA2}.
In contrast to a fully active LIS, a DMA can be used as a passive beamformer and receiver, respectively, for transmitting and receiving information.
As illustrated in Fig. {\ref{fig.01}}, when an RF signal arrives to the metalic patch of a DMA, the signals captured at each metamaterial flow through the corresponding waveguide
structure (i.e., metamaterial structure) and are forwarded to the digital signal processing unit for baseband processing.
In the transmission mode, the RF signal is generated from the RF chain
and injected into the metallic patch.
Then, the current flows through the metallic patch and is
divided into two different directions of the cylindrical post and the varactor, according to their impedance. 
Subsequently, the seperated current converges on the other side of the metallic patch and
the RF signals are transmitted through this metallic patch.
Since each metamaterial is composed of passive elements, such as resistors, capacitors, and inductors,
DMAs can inherently implement signal processing techniques such as
phase shifting and analog combining, without additional hardware cost, signal processing complexity, and energy consumption.
Hereinafter, we use the term LIS to imply a large antenna
array based on the concept of a DMA that can exploit the advantages of passive elements while
also providing active transceiver capabilities.

However, these recent works in \cite{ref.RR1,ref.RR2,ref.RR3,ref.RR4,ref.Jung2018lisul, ref.Hu2018data, ref.Hu2018assignment, ref.Jung2019reliability, ref.Hu2018positioning,ref.Hu2018hardware, ref.Han2018assisted, ref.Huang2018energy,ref.Wu2018beamforming,ref.Wu2019bfoptimization,ref.PSS00,ref.WPT00,ref.WPT01,ref.LISnew,ref.editor,ref.DMA1, ref.DMA2} have not considered the effects on spectral efficiency (SE) resulting from the use of a practical uplink frame structure in which the pilot training and data transmission period are jointly considered.
Given that statistical channel state information (CSI) is typically acquired by pilot signaling, and because the channel uses for data transmission are closely related to the length of the pilot sequence \cite{ref.Kim2018scaling},
an uplink frame structure that includes pilot training strongly impacts the achievable SE of LIS systems.
Moreover, this pilot signal will be contaminated by inter-LIS interference, similar to inter-cell interference in multi-cell MIMO environment (e.g., see \cite{ref.Hoydis2013how} and \cite{ref.Marzetta2010noncooperative}).
Therefore, accurate CSI estimations with an optimal pilot training length constitute an important challenge in multi-LIS systems where the pilot sequences are reused in adjacent LISs.
In fact, prior studies on massive MIMO \cite{ref.Kim2018scaling,ref.Hoydis2013how,ref.Marzetta2010noncooperative} do not directly apply to LIS, because the channel model of LIS is significantly different from the one used in these prior studies.
For densely located LISs, all channels will be modeled by device-specific spatially correlated Rician fading depending on the distance between each LIS and device,
however, the massive MIMO works in \cite{ref.Kim2018scaling,ref.Hoydis2013how,ref.Marzetta2010noncooperative} rely on a Rayleigh fading channel considering far-field communications.
Moreover, in LIS, each area of the large surface constitutes one of the key parameters that determine the performance of an LIS system \cite{ref.Jung2018lisul, ref.Hu2018data, ref.Hu2018assignment},
however, in existing massive MIMO works \cite{ref.Kim2018scaling,ref.Hoydis2013how,ref.Marzetta2010noncooperative}, this notion of an area is not applicable.

There has also been a number of works that look at scaling laws, such as \cite{ref.Jung2018lisul}, \cite{ref.Wu2018beamforming}, \cite{ref.scaling1}, and \cite{ref.scaling2}.
For example, recently, the authors in \cite{ref.Wu2018beamforming}, \cite{ref.scaling1}, and \cite{ref.scaling2} have analyzed the performance of an IRS-assisted system using the scaling law for a large number of reflecting elements.
Also, the work in \cite{ref.Jung2018lisul} has analyzed the performance of an LIS system based on the scaling law.
However, these recent works in \cite{ref.Jung2018lisul}, \cite{ref.Wu2018beamforming}, \cite{ref.scaling1}, and \cite{ref.scaling2} have not considered the effects of the pilot contamination on the performance perspective.
Note that large array systems such as massive MIMO have a particular operating characteristic whereby the performance is ultimately limited by the pilot contamination
based on the performance scaling law \cite{ref.Hoydis2013how}.
Hence, when we characterize the effects of a large number of reflecting elements on the performance using the scaling law,
there is a need to analyze the pilot contamination.

\subsection{Contributions}
The main contribution of this paper is an asymptotic analysis of the uplink system SE (SSE) in a multi-LIS environment that considers a practical uplink frame structure based on the 3GPP model in \cite{ref.LTE2017TR36211}.
In our model, we consider multiple LISs, each of which operates as an RF transmitter/receiver\footnote{Since an IRS is regard as a passive reflecting surface, it will not have an RF chain and thus, it is hard to estimate channels on IRS.
However, in case of an LIS operating as an RF transmitter/receiver, each LIS has its own signal processing unit and this allows the LIS to estimate CSI and detect the desired signal, as assumed in \cite{ref.Jung2018lisul, ref.Hu2018data, ref.Hu2018assignment, ref.Jung2019reliability, ref.Hu2018positioning,ref.Hu2018hardware}.}.
The SSE is typically measured as the data rates that can be simultaneously supported by a limited bandwidth in a defined geographic area \cite{ref.Miao2016mobile}.
For a given LIS that serves multiple devices, we define the SSE as the sum of the individual SE of each LIS device.
Then, we analyze the asymptotic SSE including its ergodic value, channel hardening effect, and performance bound, under pilot contamination considerations, relying on a scaling law for a large number of antennas.
The devised approximation allows for {\emph{accurate estimations of actual SSE}}, deterministically, and it also allows verifying the reliability of an LIS system.
Subsequently, we analyze the effect of pilot training in a realistic channel fading scenario in which channel states are limited by the channel coherence block and are exclusively static within limited time and frequency blocks.
The pilot training analysis provides insights on how the optimal pilot training length scales with the various parameters of LIS systems in a deterministic way.
It also reveals a particular operating characteristic of an LIS, whereby the {\emph{optimal pilot training length converges to the number of devices located within an LIS area}, as the number of antennas increases without bound.
Given the derived pilot training length, we finally derive the optimal number of devices that each LIS must schedule, to maximize the SSE.
Simulation results show that an LIS with the proposed operating parameters, including the pilot training length and the number of scheduled devices,
can achieve a maximum SSE performance both in single- and multi-LIS environments, regardless of the effect of pilot contamination and inter-LIS interference.
Moreover, the impact on an LIS system of pilot contamination can be negligible when inter-LIS interference channels are generated from spatially correlated Rayleigh fading,
which highlights a significant difference from conventional massive MIMO.

\begin{table}[]
\centering
\caption{Summary of our Notations}\vspace{-0.5cm}
\begin{tabular}{|c|c|}
\hline
\textbf{Notation}                                            & \textbf{Description}                          \\ \hline
$N$                                            & Number of LISs                         \\ \hline
$K$                                            & Number of LIS units in each LIS (equal to the number of devices)                        \\ \hline
$M$                                            & Number of antennas on each LIS unit                         \\ \hline
$L$                                            & Half length of each LIS unit                         \\ \hline
${{{\boldsymbol{h}}_{nnkk}^{\rm{L}}}}$                                            & Desired LOS channel between device $k$ and LIS unit $k$ at LIS $n$                         \\ \hline
${{{\boldsymbol{h}}_{nnjk}}}, j \ne k$                        & Intra-LIS interference channel between device $j$ and LIS unit $k$ at LIS $n$          \\ \hline
${{{\boldsymbol{h}}_{lnjk}}}, l \ne n$                       & Inter-LIS interference channel between device $j$ at LIS $l$ and LIS unit $k$ at LIS $n$ \\ \hline
${{{\boldsymbol{\hat h}}_{nnkk}^{\rm{L}}}}$                                            & Estimated channel between device $k$ and LIS unit $k$ at LIS $n$                          \\ \hline
$e_{nk}$                                                 & Channel estimation error at LIS unit $k$ on LIS $n$                             \\ \hline
$\gamma_{nk}$                           & Received SINR at LIS unit $k$ on LIS $n$                                        \\ \hline
$R_n^{\rm{SSE}}$ & Instantaneous SSE at LIS $n$                                                       \\ \hline
\end{tabular}\label{table.not}\vspace{-0.5cm}
\end{table}

The rest of this paper is organized as follows. 
Section II presents the LIS-based system model.
Section III describes the asymptotic analysis of the SSE, and
Section IV describes the performance bound and optimal pilot training length based on the results of Section III.
The optimal number of scheduled devices is also discussed in Section IV.
Simulation results are provided in Section V to support and verify the analyses, and
Section VI concludes the paper.

\textit{Notations:} Hereinafter, boldface upper- and lower-case symbols represent matrices and vectors respectively, and $\boldsymbol{I}_M$ denotes a size-$M$ identity matrix.
${\mu _X}$ and $\sigma _X^2$ denote the mean and variance of a random variable $X$, respectively.
The conjugate, transpose, and Hermitian transpose operators are denoted by ${\left(  \cdot  \right)^*}$, ${\left(  \cdot  \right)^{\rm{T}}}$, and ${\left(  \cdot  \right)^{\rm{H}}}$, respectively. 
The norm of a vector $\boldsymbol{a}$ is denoted by $\left\| {\boldsymbol{a}} \right\|$
and the Frobenious norm of a matrix $\boldsymbol{A}$ is $\| {\boldsymbol{A}} \|_{\rm{F}}$.
${\rm{E}}\left[  \cdot  \right]$, $\mathcal{O}\left(  \cdot  \right)$, $\otimes$ denote the expectation operator, big O notation, and the Kronecker product, respectively.
$\mathcal{CN}\left( {m,{\sigma ^2}} \right)$ is a complex Gaussian distribution with mean $m$ and variance $\sigma ^2$.
The important notations used in this paper are summarized in Table \ref{table.not}.

\begin{figure}
\centering
  \subfigure[]{\label{fig.1a}
  \includegraphics[width=0.48\columnwidth]{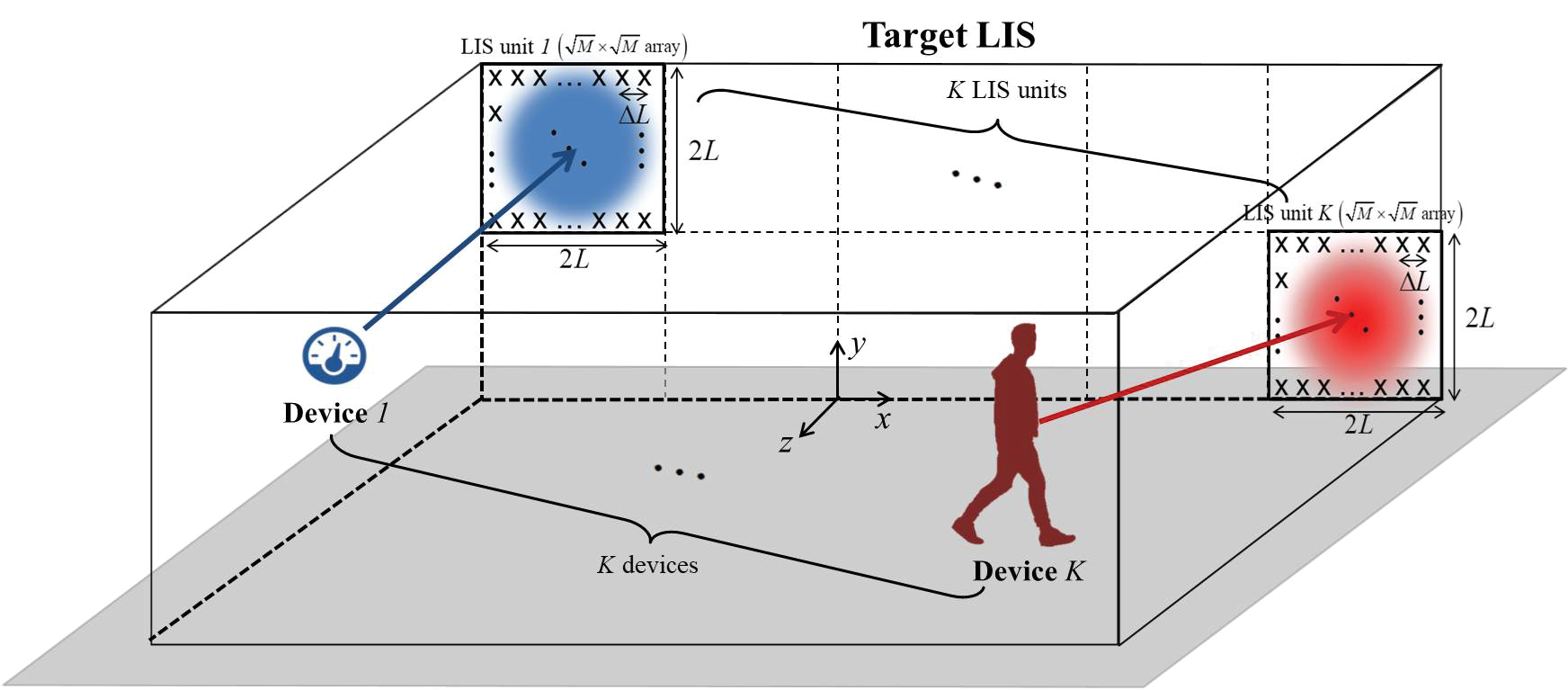}}
  \subfigure[]{\label{fig.1b}
  \includegraphics[width=0.48\columnwidth]{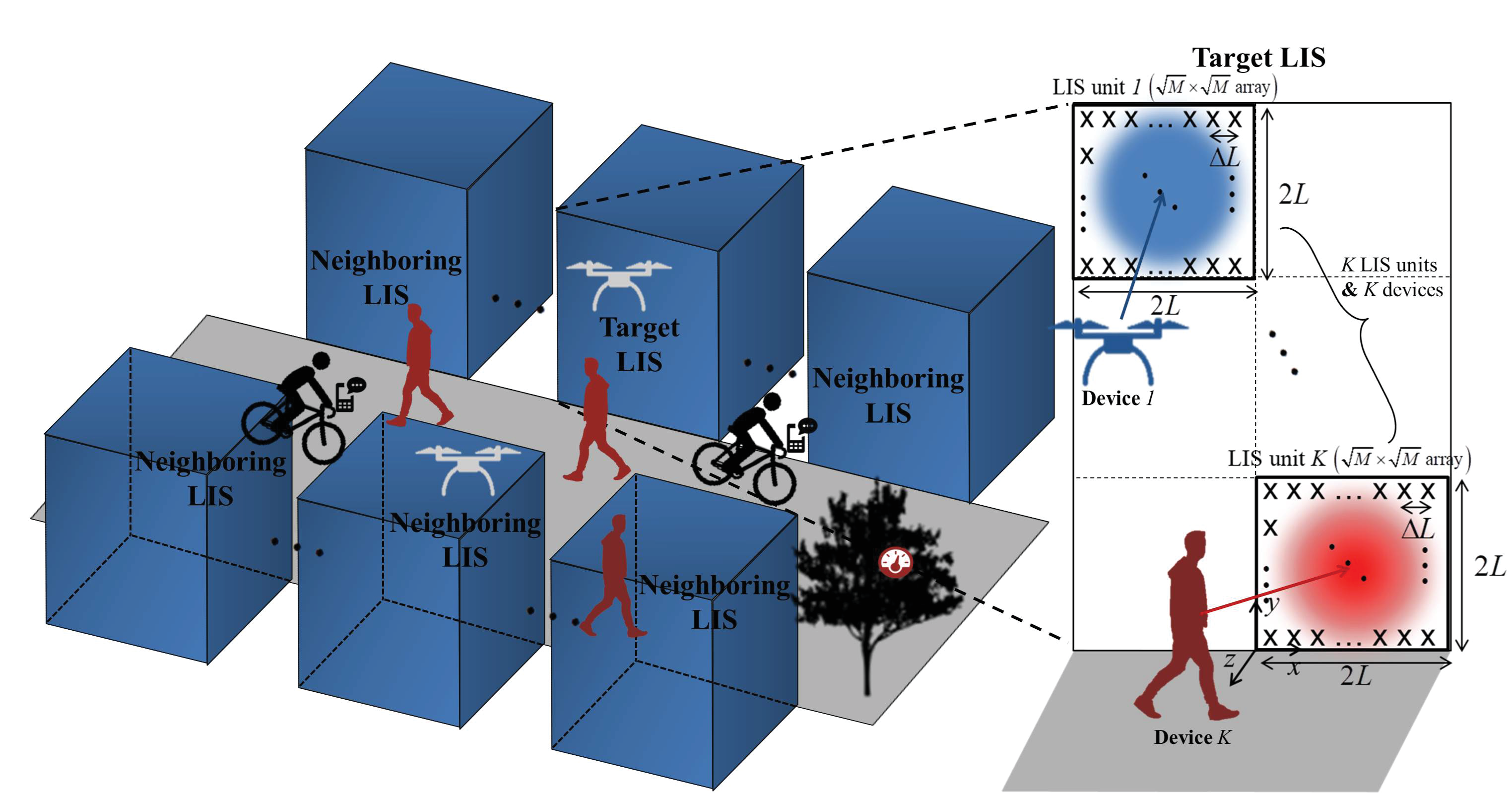}}
\caption{Illustrative system model of the uplink LIS under consideration of (a) indoor case with single LIS, (b) outdoor case with multiple LISs, with $K$ devices per each LIS.}\vspace{-0.55cm}
\label{fig.1}
\end{figure}


\section{System Model}
Consider an uplink LIS system with $N$ LISs ($N \ge 1$) sharing the same frequency band.
Recall that the term LIS is here used to denote a DMA-based large antenna array.
Each LIS is located in two-dimensional Cartesian space along the $xy$-plane, serving $K$ devices, as shown in Fig. \ref{fig.1}.
Each LIS is divided into $K$ non-overlapping LIS units, each of which serves a single-antenna device occupying a $2L \times 2L$ square shaped subarea of the entire LIS.
Therefore, we consider $K$ devices and $K$ LIS units for each LIS where device $k$ transmits its uplink signal to LIS unit $k$ (i.e., one device per LIS unit).
We assume that each LIS unit has its own signal process unit, which collects the signal received at the LIS unit
for estimating CSI and detecting uplink signal from the serving device, as assumed in \cite{ref.Hu2018assignment} and \cite{ref.Hu2018hardware}.
A large number of antennas, $M$, are deployed on the surface of each LIS unit with $\Delta L$ spacing, arranged in a rectangular lattice centered on the $(x, y)$ coordinates of the corresponding device.
Given the location $({x_{nk}},{y_{nk}},{z_{nk}})$ of device $k$ at LIS $n$, antenna $m$ of LIS unit $k$ at LIS $n$ will be located at $({x^{\rm{LIS}}_{nkm}},{y^{\rm{LIS}}_{nkm}},0)$,
and each device will transmit the uplink signal toward the center of its LIS unit.
Also, each device will control its transmission power according to a target signal-to-noise-ratio (SNR), $\gamma_{\rm{t}}$, to avoid the near-far problem.
Fig. {\ref{fig.1}} illustrates our system model for an indoor case with a single LIS and an outdoor case with multiple LISs.
In case of single LIS, as shown in Fig. \ref{fig.1a}, the desired signal is affected, exclusively, by \emph{intra-LIS interference} which is defined as the interference generated by multiple devices located within the same LIS area.
On the other hand, in case of multiple LISs, as shown in Fig. \ref{fig.1b}, the desired signal can be affected by both \emph{intra-LIS} and \emph{inter-LIS interference} simultaneously.
Here, \emph{inter-LIS interference} corresponds to the interference generated by devices serviced by other LISs.


\subsection{Wireless Channel Model}
In LIS systems, entire man-made structures are electromagnetically active and can be used for wireless communication.
We then consider the LIS channel ${\boldsymbol{h}}_{nnkk}^{\rm{L}} \in \mathbb{C}{^M}$ between device $k$ at LIS $n$ and LIS unit $k$ part of LIS $n$ as a LOS path defined by:
\begin{equation}
{\boldsymbol{h}}_{nnkk}^{\rm{L}}  = {\left[ {\beta _{nnkk1}^{\rm{L}}{h_{nnkk1}}, \cdots ,\beta _{nnkkM}^{\rm{L}}{h_{nnkkM}}} \right]^{\rm{T}}}, \label{eq.1}
\end{equation}
where $\beta _{nnkkm}^{\rm{L}} = \alpha _{nnkkm}^{\rm{L}}l_{nnkkm}^{\rm{L}}$ and ${h_{nnkkm}} = \exp \left( { - j2\pi {d_{nnkkm}}/\lambda } \right)$ denote a LOS channel gain and state, respectively, between device $k$ at LIS $n$ and antenna $m$ of LIS unit $k$ part of LIS $n$ \cite{ref.Tse2005fundamentals}.
The terms $\alpha_{nnkkm}^{\rm{L}} = \sqrt { z_{nk}/d_{nnkkm}} $ and $l_{nnkkm}^{\rm{L}} = 1/\sqrt {4\pi d_{nnkkm}^2}$ represent, respectively, the antenna gain and free space path loss attenuation,
where ${d_{nnkkm}}$ is the distance between device $k$ at LIS $n$ and antenna $m$ of LIS unit $k$ part of LIS $n$ as follows:
\begin{equation}
{d_{nnkkm}} = \sqrt {{{( {{x_{nk}} - {{x}^{\rm{LIS}}_{nkm}}} )}^2} + {{( {{y_{nk}} - {{y}^{\rm{LIS}}_{nkm}}} )}^2} + z_{nk}^2},
\end{equation}
and $\lambda$ is the wavelength of a signal.
We model the interference channel ${{\boldsymbol{h}}_{lnjk}} \in \mathbb{C}{^M}$ between device $j$ at LIS $l$ and LIS unit $k$ part of LIS $n$ as a Rician fading channel with Rician factor $\kappa _{lnjk}$, given by:
\begin{equation}
{{\boldsymbol{h}}_{lnjk}}
=\sqrt {\frac{{{\kappa _{lnjk}}}}{{{\kappa _{lnjk}} + 1}}} {\boldsymbol{h}}_{lnjk}^{\rm{L}} + \sqrt {\frac{1}{{{\kappa _{lnjk}} + 1}}} {\boldsymbol{h}}_{lnjk}^{{\rm{NL}}}, \label{eq.2}
\end{equation}
where ${\boldsymbol{h}}_{lnjk}^{\rm{L}} \in \mathbb{C}{^M} ={\left[ {\beta _{lnjk1}^{\rm{L}}{h_{lnjk1}}, \cdots ,\beta _{lnjkM}^{\rm{L}}{h_{lnjkM}}} \right]^{\rm{T}}}$ and ${\boldsymbol{h}}_{lnjk}^{\rm{NL}} \in \mathbb{C}{^M} =  {\boldsymbol{R}}_{lnjk}^{1/2}{{\boldsymbol{g}}_{lnjk}}$ denote the deterministic LOS and the correlated NLOS component, respectively.
For notational simplicity, we define ${{\bar{\boldsymbol{h}}}}_{lnjk}=\sqrt {\frac{{{\kappa _{lnjk}}}}{{{\kappa _{lnjk}} + 1}}} {\boldsymbol{h}}_{lnjk}^{\rm{L}}$ and ${{\tilde{\boldsymbol{h}}}}_{lnjk}=\sqrt {\frac{1}{{{\kappa _{lnjk}} + 1}}} {\boldsymbol{h}}_{lnjk}^{{\rm{NL}}}$.
Here, if $l=n$ and $j\ne k$, then ${{\boldsymbol{h}}_{lnjk}}$ indicates the intra-LIS interference channel,
otherwise, if $l\ne n$ $\forall j,k$, then ${{\boldsymbol{h}}_{lnjk}}$ indicates the inter-LIS interference channel.
\begin{figure}[]
\centering
\includegraphics[width=0.35\columnwidth] {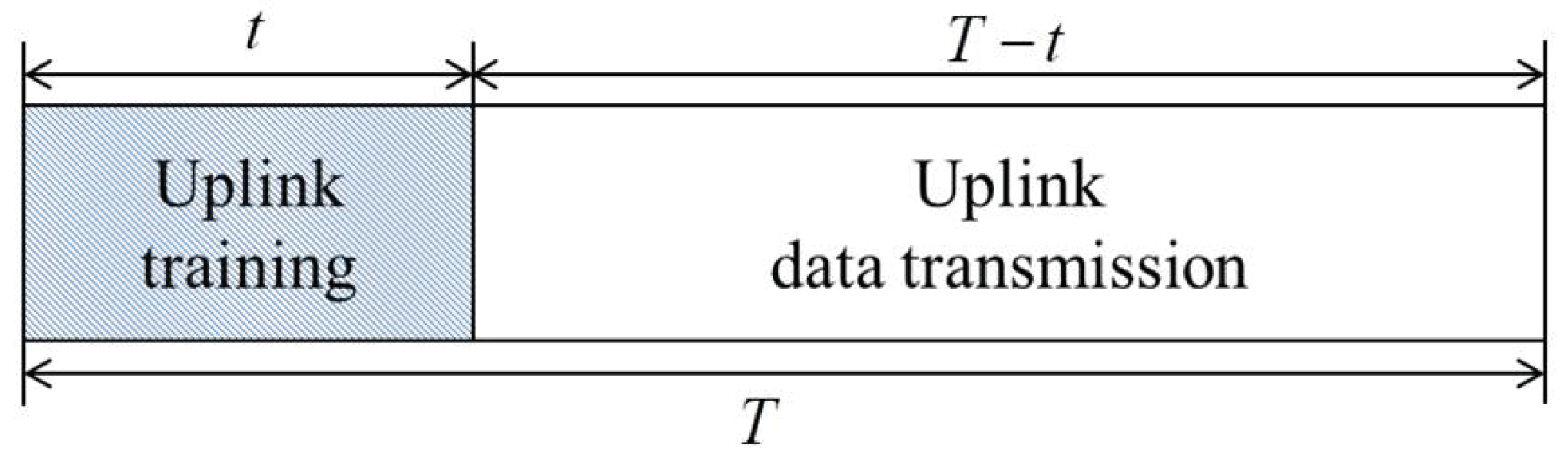}\vspace{-0.4cm}
\caption{Illustrative uplink frame structure with a pilot training period $t$ and a data transmission period $T-t$.}\vspace{-0.5cm}
\label{fig.2}
\end{figure}
Considering $P$ dominant paths among all NLOS paths, we define ${{\boldsymbol{R}}_{lnjk}} \in \mathbb{C}{^{M \times P}}$ and ${{\boldsymbol{g}}_{lnjk}} = {\left[ {{g_{lnjk1}}, \cdots ,{g_{lnjkP}}} \right]^{\rm{T}}} \sim \mathcal{CN}\left( {{\boldsymbol{0}},{{\boldsymbol{I}}_P}} \right)$ to be the deterministic correlation matrix and an independent fast-fading channel vector between device $j$ at LIS $l$ and LIS unit $k$ part of LIS $n$, respectively.
Since the LIS is deployed on the horizontal plane, as shown in Fig. \ref{fig.1},
we can model it as a uniform planar array \cite{ref.Song2017UPA}.
Then, the correlation matrix can be defined as
${\boldsymbol{R}}_{lnjk}^{1/2} = {\boldsymbol{l}}_{lnjk}^{{\rm{NL}}}{{\boldsymbol{D}}_{lnjk}} $, where 
${\boldsymbol{l}}_{lnjk}^{{\rm{NL}}} = {\rm{diag}}\left( {l_{lnjk1}^{{\rm{NL}}}  \cdots ,l_{lnjkM}^{{\rm{NL}}}} \right)$ is a diagonal matrix that includes the path loss attenuation factors $l_{lnjkm}^{{\rm{NL}}} = d_{lnjkm}^{ - {\beta _{{\rm{PL}}}}/2}$ with a path loss exponent ${\beta _{{\rm{PL}}}}$ and
${{\boldsymbol{D}}_{lnjk}} = \left[ {\alpha _{lnjk1}^{{\rm{NL}}}{\boldsymbol{d}}\left( {\phi _{lnjk1}^{\rm{v}},\phi _{lnjk1}^{\rm{h}}} \right), \cdots ,\alpha _{lnjkP}^{{\rm{NL}}}{\boldsymbol{d}}\left( {\phi _{lnjkP}^{\rm{v}},\phi _{lnjkP}^{\rm{h}}} \right)} \right]$.
${\boldsymbol{d}}\left( {\phi _{lnjkp}^{\rm{v}},\phi _{lnjkp}^{\rm{h}}} \right) \in \mathbb{C}{^M}$ is the NLOS path $p$ at given angles of $\left( {\phi _{lnjkp}^{\rm{v}},\phi _{lnjkp}^{\rm{h}}} \right)$ defined as:
\begin{gather}
{\boldsymbol{d}}\left( {\phi _{lnjkp}^{\rm{v}},\phi _{lnjkp}^{\rm{h}}} \right) = \frac{1}{{\sqrt M }}{{\boldsymbol{d}}_{\rm{v}}}\left( {\phi _{lnjkp}^{\rm{v}}} \right) \otimes {{\boldsymbol{d}}_{\rm{h}}}\left( {\phi _{lnjkp}^{\rm{h}}} \right), \label{eq.DNLOS}\\
{{\boldsymbol{d}}_{\rm{v}}}\left( {\phi _{lnjkp}^{\rm{v}}} \right) = {\left[ {1,{\rm{ }}{e^{j\frac{{2\pi \Delta L}}{\lambda }\phi _{lnjkp}^{\rm{v}}}}, \cdots ,{\rm{ }}{e^{j\frac{{2\pi \Delta L}}{\lambda }\left( {\sqrt M  - 1} \right)\phi _{lnjkp}^{\rm{v}}}}} \right]^{\rm{T}}}, \\
{{\boldsymbol{d}}_{\rm{h}}}\left( {\phi _{lnjkp}^{\rm{h}}} \right) = {\left[ {1,{\rm{ }}{e^{j\frac{{2\pi \Delta L}}{\lambda }\phi _{lnjkp}^{\rm{h}}}}, \cdots ,{\rm{ }}{e^{j\frac{{2\pi \Delta L}}{\lambda }\left( {\sqrt M  - 1} \right)\phi _{lnjkp}^{\rm{h}}}}} \right]^{\rm{T}}}, 
\end{gather}
where $\phi _{lnjkp}^{\rm{v}} = \sin \theta _{lnjkp}^{\rm{v}}$ and $\phi _{lnjkp}^{\rm{h}} = \sin \theta _{lnjkp}^{\rm{h}}\cos \theta _{lnjkp}^{\rm{v}}$ when the elevation and azimuth angles of path $p$ between device $j$ at LIS $l$ and LIS unit $k$ part of LIS $n$ are $\theta _{lnjkp}^{\rm{v}}$ and $\theta _{lnjkp}^{\rm{h}}$, respectively \cite{ref.Han2014design}. 
Further, $\alpha _{lnjkp}^{{\rm{NL}}} = \sqrt {\cos \theta _{lnjkp}^{\rm{v}}\cos \theta _{lnjkp}^{\rm{h}}}$ denotes the antenna gain of path $p$ with ${\theta _{lnjkp}} \in \left[ { - \frac{\pi }{2},\frac{\pi }{2}} \right]$ and ${\theta _{lnjkp}} \in \left\{ {\theta _{lnjkp}^{\rm{v}},\theta _{lnjkp}^{\rm{h}}} \right\}$.

\subsection{Uplink Pilot Training}
We consider that an MF is used at the LIS to amplify the desired signals and suppress interfering signals.
This MF receiver requires CSI which can be estimated by pilot signaling with known pilot signals being transmitted from the device to the LIS.
The device transmits its data signals immediately after sending the pilot signals within the channel coherence time $T$ in which the uplink channel is approximately static.
We consider the uplink frame structure shown in Fig. \ref{fig.2}, in which the total duration of $T$ channel uses is divided into a $t$ period used for pilot training and a $T-t$ period used for data transmission.
Every device simultaneously transmits $t\ge {K}$ orthogonal pilot sequences over the uplink channel to the LIS, so that the required CSI can be acquired.
Given that those ${K}$ pilot sequences are pairwise orthogonal to each other, we have ${\boldsymbol{\Psi }}^{\rm{H}}{\boldsymbol{\Psi }} = {{\boldsymbol{I}}_{K}}$, where ${\boldsymbol{\Psi }} = [{{\boldsymbol{\psi}} _1}, ... ,{{\boldsymbol{\psi}} _{K}}]$
and ${{\boldsymbol{\psi}} _k}$ is the $t \times 1$ pilot sequence for device $k$.

For the multi-LIS scenario in which the same frequency band is shared by all LISs and adjacent LISs reuse the pilot sequences,
the pilot symbols between adjacent LISs are no longer orthogonal to each other and this non-orthogonality causes \emph{pilot contamination}.
In large antenna-array systems such as massive MIMO and LIS, the performance can be dominantly limited by residual interference from pilot contamination as explained in \cite{ref.Kim2018scaling,ref.Hoydis2013how,ref.Marzetta2010noncooperative}.
Since LISs will be located more densely than BSs, the LIS channels associated with pilot contamination will be significantly different than those of massive MIMO,
and hence, prior studies on pilot contamination for massive MIMO \cite{ref.Kim2018scaling,ref.Hoydis2013how,ref.Marzetta2010noncooperative} do not directly apply to LIS.
In order to verify the effect of pilot contamination in an LIS system theoretically, we consider such multi-LIS scenario
in which a total of $N$ LISs share the same frequency band and each LIS reuses $K$ pilot sequences.
Moreover, all LISs are assumed to use the same uplink frame structure shown in Fig. \ref{fig.2}, whereby a pilot sequence $k$ is allocated to device $k$ for all $1 \le k \le {K}$.
The uplink pilot sequence received from all devices at LIS unit $k$ part of LIS $n$ during period $t$ will be:
\begin{equation}
{\boldsymbol{Y}}_{nk}^{\rm{p}} = \sqrt {t{\rho _{{{\rm{p}}_{nk}}}}} {\boldsymbol{h}}_{nnkk}^{\rm{L}}{\boldsymbol{\psi }}_k^{\rm{H}} + \sum\nolimits_{j \ne k}^{K} {\sqrt {t{\rho _{{\rm{p}}_{nj}}}} {\boldsymbol{h}}_{nnjk}{\boldsymbol{\psi }}_j^{\rm{H}}}  + \sum\nolimits_{l \ne n}^{N} {\sum\nolimits_{j = 1}^{K} {\sqrt {t{\rho _{{\rm{p}}_{lj}}}} {\boldsymbol{h}}_{lnjk}{\boldsymbol{\psi }}_j^{\rm{H}}} }  + {{\boldsymbol{N}}_{nk}},\label{eq.yp}
\end{equation}
where ${\rho _{{{\rm{p}}_{nk}}}}$, ${\rho _{{{\rm{p}}_{nj}}}}$, and ${\rho _{{{\rm{p}}_{lj}}}}$ are the transmit SNRs for the pilot symbols of device $k$ at LIS $n$, device $j$ at LIS $n$, device $j$ at LIS $l$, respectively, and ${\boldsymbol{N}}_{nk} \in \mathbb{C}{^{M \times t}}\sim \mathcal{CN}\left( {{\boldsymbol{0}},{{\boldsymbol{I}}_M}} \right)$ is a noise matrix at LIS unit $k$ part of LIS $n$.
We assume that the target SNR for a pilot symbol is assumed to be equal to $\rho_{\rm{p}}$
and each device controls its pilot power toward the center of the corresponding LIS unit.
On the basis of orthogonal characteristic of the pilot sequences, each LIS unit $k$ multiplies the received pilot signal by ${\boldsymbol{\psi }}_k$ for channel estimation.
After multiplying ${\boldsymbol{\psi }}_k$ at both sides of (\ref{eq.yp}), we have
\begin{equation}
{\boldsymbol{Y}}_{nk}^{\rm{p}}{{\boldsymbol{\psi }}_k} 
= \sqrt {t{\rho _{{\rm{p}}_{nk}}}} {{\boldsymbol{h}}_{nnkk}^{\rm{L}}} + \sum\nolimits_{l \ne n}^{N} {\sqrt {t{\rho _{{{\rm{p}}_{lk}}}}} {{{\boldsymbol{ h}}}_{lnkk}}}  + {{\boldsymbol{N}}_{nk}}{{\boldsymbol{\psi }}_k}.
\end{equation}
In most prior research on pilot contamination in large antenna-array systems such as in \cite{ref.Hoydis2013how} and \cite{ref.Jose2011PC}, the minimum mean square error (MMSE) channel estimator is assumed to estimate a desired channel given that the BS has knowledge of every correlation matrix between itself and interfering users located in adjacent cells.
However, this assumption is impractical for LIS systems because a massive number of devices will be connected to an LIS,
and thus, processing complexity will increase tremendously when estimating and sharing device information.
Therefore, we consider a simple least square (LS) estimator which does not require such information as a practical alternative \cite{ref.Khansefid2015LS}. 
The LS estimate of the deterministic desired channel ${{\boldsymbol{h}}_{nnkk}^{\rm{L}}}$ is then obtained as:
\begin{equation}
{{{\boldsymbol{\hat h}}}_{nnkk}} = {{\boldsymbol{h}}_{nnkk}^{\rm{L}}} + {{\boldsymbol{e}}_{nk}},\label{eq.h_hat}
\end{equation}
where ${{\boldsymbol{e}}_{nk}}$ indicates the estimation error vector given by ${{\boldsymbol{e}}_{nk}}= \sum\nolimits_{l \ne n}^{N} {\sqrt {\frac{{{\rho _{{{\rm{p}}_{lk}}}}}}{{{\rho _{{\rm{p}}_{nk}}}}}} {{{\boldsymbol{ h}}}_{lnkk}}}  + \frac{1}{{\sqrt {t{\rho _{{\rm{p}}_{nk}}}} }}{{\boldsymbol{w}}_{nk}}$
and ${{\boldsymbol{w}}_{nk}}=\left[ w_{nk1},\cdots,w_{nkM}\right]^{\rm{T}}\in \mathbb{C}{^M} \sim \mathcal{CN} \left( {{\boldsymbol{0}},{{\boldsymbol{I}}_M}} \right)$.


\subsection{Instantaneous uplink SSE}
The uplink signal received from all devices at LIS unit $k$ part of LIS $n$ is given by:
\begin{equation}
{\boldsymbol{y}}_{nk} = \sqrt {{\rho _{{{nk}}}}} {\boldsymbol{h}}_{nnkk}^{\rm{L}}{x_{nk}} + \sum\limits_{j \ne k}^{K} {\sqrt {{\rho _{{nj}}}} {\boldsymbol{h}}_{nnjk}{x_{nj}}}  + \sum\limits_{l \ne n}^{N} {\sum\limits_{j = 1}^{K} {\sqrt {{\rho _{{lj}}}} {\boldsymbol{h}}_{lnjk}{x_{lj}}} }  + {{\boldsymbol{n}}_{nk}},
\end{equation}
where ${x_{nk}}$, ${x_{nj}}$, and ${x_{lj}}$ are uplink transmit signals of device $k$ at LIS $n$, device $j$ at LIS $n$, and device $j$ at LIS $l$, respectively, 
and ${\rho_{nk}}$, ${\rho_{nj}}$, and ${\rho_{lj}}$ are their transmit SNRs.
Also, ${{\boldsymbol{n}}_{nk}} \in\mathbb{C} {^M} \sim \mathcal{CN}\left( {{\boldsymbol{0}},{{\boldsymbol{I}}_M}} \right)$ is the noise vector at LIS unit $k$ part of LIS $n$.
Given a linear receiver ${\boldsymbol{f}}_{nk}^{\rm{H}}$ resulting from the waveguide structures for signal detection, we will have
\begin{equation}
{\boldsymbol{f}}_{nk}^{\rm{H}}{\boldsymbol{y}}_{nk} = \sqrt {{\rho _{{{nk}}}}} {\boldsymbol{f}}_{nk}^{\rm{H}}{\boldsymbol{h}}_{nnkk}^{\rm{L}}{x_{nk}} + \sum\limits_{j \ne k}^{K} {\sqrt {{\rho _{{nj}}}}{\boldsymbol{f}}_{nk}^{\rm{H}}{\boldsymbol{h}}_{nnjk}{x_{nj}}}  + \sum\limits_{l \ne n}^{N} {\sum\limits_{j = 1}^{K} {\sqrt {{\rho _{{lj}}}} {\boldsymbol{f}}_{nk}^{\rm{H}}{\boldsymbol{h}}_{lnjk}{x_{lj}}} }  + {\boldsymbol{f}}_{nk}^{\rm{H}}{{\boldsymbol{n}}_{nk}}.
\end{equation}
Since the propagation amplitude in $[0,1]$ resulting from the waveguide structure is achievable
by changing the values of resistors in each metamaterial element \cite{ref.R2A4},
we consider an MF receiver such that ${{\boldsymbol{f}}_{nk}} =\left[ f_{nk1},\cdots,f_{nkM}\right]^{\rm{T}}= {{\boldsymbol{\hat h}}_{nnkk}}/||{{\boldsymbol{\hat h}}_{nnkk}}||$
where $|{{{f}}_{nkm}}|\in [0,1], \forall m$.
Under the imperfect CSI results from an LS estimator, ${{\boldsymbol{\hat h}}_{nnkk}}$ can be obtained from (\ref{eq.h_hat}) where ${{\boldsymbol{e}}_{nk}}$ is the estimation error vector uncorrelated with ${{\boldsymbol{n}}}_{nk}$ \cite{ref.Poor1994introduction}.
Therefore, the received signal-to-interference-plus-noise ratio (SINR) at LIS unit $k$ part of LIS $n$ will be:
\begin{equation}
{\gamma _{nk}} = \frac{{{\rho _{nk}}{{\left\| {{{\boldsymbol{h}}_{nnkk}^{\rm{L}}}} \right\|}^4}}}{{{\rho _{nk}}{{\left| {{\boldsymbol{e}}_{nk}^{\rm{H}}{{\boldsymbol{h}}_{nnkk}^{\rm{L}}}} \right|}^2} + \sum\limits_{j \ne k}^{K} {{\rho _{nj}}{{\left| {{\boldsymbol{\hat h}}_{nnkk}^{\rm{H}}{{\boldsymbol{h}}_{nnjk}}} \right|}^2}}  + \sum\limits_{l \ne n}^N {\sum\limits_{j = 1}^{K} {{\rho _{lj}}{{\left| {{\boldsymbol{\hat h}}_{nnkk}^{\rm{H}}{{\boldsymbol{h}}_{lnjk}}} \right|}^2} + {{\left\| {\big({\boldsymbol{h}}_{nnkk}^{\rm{L}}\big)^{\rm{H}} + {\boldsymbol{e}}_{nk}^{\rm{H}}} \right\|}^2}} } }}.\label{eq.gamma}\\
\end{equation}
For notational simplicity, we define
${\gamma _{nk}}= {{{\rho _{nk}}{S_{nk}}}}/{{{I_{nk}}}}$, where ${S_{nk}} = {{\left\| {{\boldsymbol{h}}_{nnkk}^{\rm{L}}} \right\|}^4}$ and
\begin{equation}
{I_{nk}} = {{{\rho _{nk}}X_{nk} + \sum\nolimits_{j \ne k}^{K} {{\rho _{nj}}Y_{njk}}}  + \sum\nolimits_{l \ne n}^N {\sum\nolimits_{j = 1}^{K} {{\rho _{lj}}Y_{lnjk} + Z_{nk}} } }, \label{eq.Ink}
\end{equation}
where ${X_{nk}} {=} {{\left| {{\boldsymbol{e}}_{nk}^{\rm{H}}{{\boldsymbol{h}}_{nnkk}^{\rm{L}}}} \right|}^2},$ 
${Y_{njk}} {=} {{\left| {{\boldsymbol{\hat h}}_{nnkk}^{\rm{H}}{{\boldsymbol{h}}_{nnjk}}} \right|}^2},$
${Y_{lnjk}} {=} {{\left| {{\boldsymbol{\hat h}}_{nnkk}^{\rm{H}}{{\boldsymbol{h}}_{lnjk}}} \right|}^2},$
and ${Z_{nk}} {=} {{\left\| \big({{\boldsymbol{h}}_{nnkk}^{\rm{L}}\big)^{\rm{H}} {+} {\boldsymbol{e}}_{nk}^{\rm{H}}} \right\|}^2}.$ 
Here, we note that the SINR of an LIS will differ from a classical massive MIMO SINR. 
For instance, in the massive MIMO case, the desired signal power is derived as $S_{nk} = {{\| {{{\boldsymbol{\hat h}}_{nnkk}}} \|}^4}$
since the BS only has knowledge about the estimated channel ${{{\boldsymbol{\hat h}}_{nnkk}}}$.
However, in the considered LIS system, ${{\left\| {{{\boldsymbol{h}}_{nnkk}^{\rm{L}}}} \right\|}^4}$ can be known at the LIS because it is a deterministic value obtained by $\big(\sum\nolimits_{m=1}^M({\beta _{nnkkm}^{\rm{L}}})^2\big)^2$
and the LIS can know $\sum\nolimits_{m}({\beta _{nnkkm}^{\rm{L}}})^2$ based on the channel quality information (CQI) feedback from device $k$.
Therefore, we have $S_{nk} = {{\| {{{\boldsymbol{h}}_{nnkk}}} \|}^4}$ in the considered LIS system,
and this is a key difference between an LIS and a massive MIMO in terms of their SINR expression.

Considering $t$ and $T-t$ periods used for pilot training and data transmission, respectively,
the instantaneous SSE can be obtained as follows:
\begin{equation}
{R_n^{{\rm{SSE}}}} =\left( {1 - \frac{t}{T}} \right)\sum\limits_{k = 1}^{K} {{R_{nk}}}
=\left( {1 - \frac{t}{T}} \right)\sum\limits_{k = 1}^{K} {\log \left( {1 + {\gamma _{nk}}} \right). } \label{eq.RSSE}
\end{equation}
Given this instantaneous SSE, we will be able to analyze the asymptotic value of the SSE
and devise an optimal pilot training length $t$ that maximizes the asymptotic SSE as $M$ increases without bound.
Note that in the following sections, we use a generalized value of $N \ge 1$ in order to analyze both single- and multi-LIS cases, simultaneously (i.e., $N=1$ and $N\ge2$ indicate a single-LIS 
and $N$-LIS cases, respectively).

\section{Asymptotic SSE Analysis} 
We analyze the asymptotic value of the SSE under consideration of the pilot contamination as $M$ increases to infinity. 
In an uplink LIS system with an MF receiver, as $M$ goes to infinity, the desired signal power will be:
${S_{nk}} = \mathop {\lim }\limits_{M \to \infty } {\left( {\sum\nolimits_{m}^M {{{\left( {\beta _{nnkkm}^{\rm{L}}} \right)}^2}} } \right)^2}.$
Based on the scaling law for $M$, $S_{nk}$ is calculated by the squared sum of positive $M$ elements
and increases with $\mathcal{O}\left( M^2 \right)$.
Hence, $S_{nk}/M^2$ converges to a limited, deterministic value as $M$ goes to infinity.
We then have 
${\gamma_{nk}} - {{\bar \gamma}_{nk}} \xrightarrow[M \to \infty ]{} 0$, where
\begin{equation}
{{\bar \gamma} _{nk}} = \frac{{{\rho _{nk}}p_{nk}^2}}{{I_{nk}/M^2}},\label{eq.gamma_nk2}
\end{equation}
where $p_{nk}={ {\sum\nolimits_{m}^M {{{\left( {\beta _{nnkkm}^{\rm{L}}} \right)}^2 /M}} } }$.
We can observe from (\ref{eq.gamma_nk2}) that the distribution of ${{\bar\gamma} _{nk}}$ depends exclusively on the distribution of $I_{nk}$.
In order to analyze the distribution of $I_{nk}$ theoretically, we derive the following lemmas.
Here, we define $p$-th column vector and $m$-th row vector of ${\boldsymbol{R}}_{lnjk}^{1/2}$ as ${\boldsymbol{c}}_{lnjkp}$ and ${\boldsymbol{r}}_{lnjkm}$, respectively,
such that ${\boldsymbol{R}}_{lnjk}^{1/2} = \left[{{\boldsymbol{c}}_{lnjk1},\cdots,{\boldsymbol{c}}_{lnjkP}}\right]= \left[{{\boldsymbol{r}}_{lnjk1}^{\rm{H}},\cdots,{\boldsymbol{r}}_{lnjkM}^{\rm{H}}}\right]^{\rm{H}}$,
where ${\boldsymbol{c}}_{lnjkp} \in \mathbb{C}{^{M\times 1} }$ and ${\boldsymbol{r}}_{lnjkm} \in \mathbb{C}{^{1 \times P}}$. 

{\bf{{Lemma 1.}}} The mean of $X_{nk}$ is obtained by $ {{\mu} _{{X_{nk}}}}=\sigma _{{x_{nk}}}^2 + \left| {{\mu _{{x_{nk}}}}}\right|^2$, where
\begin{align}
{{\mu _{{x_{nk}}}}} &= \sum\limits_{l \ne n}^{N} {\sqrt {\frac{{{\rho _{{{\rm{p}}_{lk}}}}}}{{{\rho _{{\rm{p}}_{nk}}}}}}  {{{{\boldsymbol{\bar h}}}_{lnkk}^{\rm{H}}}}{{{\boldsymbol{h}}_{nnkk}^{\rm{L}}}} }, \\
\sigma _{{x_{nk}}}^2  &= \sum\limits_{l \ne n}^{N} {\frac{{{\rho_{{\rm{p}}_{lk}}}}}{{{\rho _{{{\rm{p}}_{nk}}}}\left(\kappa _{lnkk}+1\right)}}} \sum\limits_{p = 1}^P {\left| {{\boldsymbol{c}}_{lnkkp}^{\rm{H}}{{\boldsymbol{h}}_{nnkk}^{\rm{L}}}} \right|}^2  + \frac{1}{{t{\rho _{{{\rm{p}}_{nk}}}}}}\sum\limits_{m = 1}^M \left({\beta _{nnkkm}^{\rm{L}}}\right)^2. 
\end{align}
\begin{proof}
The detailed proof is presented in Appendix A.
\end{proof}

{\bf{{Lemma 2.}}} The mean values of $Y_{njk}$ and $Y_{lnjk}$ follow ${\mu _{{Y_{njk}}}} -  {{\bar\mu} _{{Y_{njk}}}}\xrightarrow[M \to \infty]{}0$ and ${\mu _{{Y_{lnjk}}}} - {\bar \mu _{{Y_{lnjk}}}}\xrightarrow[M \to \infty]{} 0$, respectively, 
where ${\bar \mu _{{Y_{njk}}}}$ and ${\bar \mu _{{Y_{lnjk}}}}$ are given, respectively, in (\ref{eq.B.MYnjk}) and (\ref{eq.B.MYlnjk}).
\begin{proof}
The detailed proof is presented in Appendix B.
\end{proof}
%

{\bf{{Lemma 3.}}} The mean of $Z_{nk}$ is obtained by
${{\mu} _{{Z_{nk}}}}=\sum\nolimits_{m=1}^M\left({\sigma _{{z_{nkm}}}^2 + \left| {{\mu _{{z_{nkm}}}}}\right|^2}\right)
$, where
\begin{align}
\mu _{{z_{nkm}}} &= \beta _{nnkkm}^{\rm{L}}{h_{nnkkm}^*} + \sum\limits_{l \ne n}^{N} {\sqrt {\frac{{{\rho _{{{\rm{p}}_{lk}}}}{\kappa _{lnkk}}}}{{\rho _{{{\rm{p}}_{nk}}}}\left({{\kappa _{lnkk}} + 1}\right)}} \beta _{lnkkm}^{\rm{L}}{h_{lnkkm}^*}},\\
\sigma _{{z_{nkm}}}^2  &= \sum\limits_{l \ne n}^{N} {\frac{{{\rho_{{\rm{p}}_{lk}}}{\left\| {\boldsymbol{r}}_{lnkkm} \right\|}^2 }}{{{\rho _{{{\rm{p}}_{nk}}}}\left({{{\kappa _{lnkk}} + 1}}\right)}}}  + \frac{1}{{t{\rho _{{{\rm{p}}_{nk}}}}}}. 
\end{align}
\begin{proof}
The detailed proof is presented in Appendix C.
\end{proof}
In Lemmas 1--3, the variables, $ {{\mu} _{{X_{nk}}}}$, $ {{\bar\mu} _{{Y_{njk}}}}$, $ {{\bar\mu} _{{Y_{lnjk}}}}$, and $ {{\mu} _{{Z_{nk}}}}$, are obtained by the deterministic information such as the locations of the devices and covariance matrices.
On the basis of Lemmas 1--3, we can asymptotically derive the mean of $I_{nk}$ from (\ref{eq.Ink}) as follows:
\begin{equation}
{\mu _{{I_{nk}}}} -  {{\bar\mu} _{{I_{nk}}}}\xrightarrow[\hspace{-2pt}M \to \infty \hspace{-2pt}]{}0,
\end{equation}
where
\begin{equation}
{{\bar\mu} _{{I_{nk}}}}= {{{\rho _{nk}}{{\mu}}_{X_{nk}} + \sum\limits_{j \ne k}^{K} {{\rho _{nj}}{\bar{\mu}}_{Y_{njk}} }}  + \sum\limits_{l \ne n}^N {\sum\limits_{j = 1}^{K} {{\rho _{lj}}{\bar{\mu}}_{Y_{lnjk}}  + {{\mu}}_{Z_{nk}} } } }.\label{eq.MInk}
\end{equation}
Since the variance of ${\bar \gamma}_{nk}$ exclusively depends on the variance of $I_{nk}/M^2$ from (\ref{eq.gamma_nk2}), the following Lemma 4 is used to obtain ${{\sigma}}_{I_{nk}}^2/M^4$ based on the scaling law for $M$.

{\bf{{Lemma 4.}}} According to the scaling law for $M$, the variance of $I_{nk}/M^2$ asymptotically follows
${\sigma _{{I_{nk}}}^2}/{M^4}\xrightarrow[M \to \infty]{\rm{}}0.$
\begin{proof}
The detailed proof is presented in Appendix D.
\end{proof}

Lemma 4 shows that $I_{nk}/M^2$ converges to the deterministic value ${{\bar\mu} _{{I_{nk}}}}/M^2$ without any variance, as $M$ increases.
Then, ${{\bar \gamma} _{nk}}$ converges to a deterministic value as $M$ increases,
and finally, we have the following Theorem 1 related to the asymptotic convergence of $R_n^{\rm{SSE}}$.

{\bf{{Theorem 1.}}} As $M$ increases to infinity, we have the following asymptotic convergence of SSE:
$R_n^{\rm{SSE}} - {\bar \mu}_n^{\rm{SSE}} \xrightarrow[M \to \infty ]{} 0, $
where
\begin{equation}
{\bar \mu}_n^{\rm{SSE}}= \left( {1 - \frac{t}{T}} \right)\sum\limits_{k = 1}^{K} {\log \left( {1 + \frac{{M^2{\rho _{nk}}p_{nk}^2}}{{{\bar \mu}_{I_{nk}}}}} \right)}.\label{eq.MSSE}
\end{equation}
\begin{proof}
The detailed proof is presented in Appendix E.
\end{proof}

Theorem 1 shows that the multi-LIS system will experience a channel hardening effect resulting in the deterministic SSE.
This deterministic SSE provides the improved system reliability and a low latency.
Moreover, we can observe from (\ref{eq.MInk}) and (\ref{eq.MSSE}) that the asymptotic SSE can be obtained from the deterministic information such as the locations of the devices and correlation matrices.
Therefore, this asymptotic approximation enables accurate estimation of the SSE without the need for extensive simulations.
Next, we use this asymptotic SSE to derive a performance bound on the SSE, asymptotically, by analyzing ${\bar \mu}_n^{\rm{SSE}}$ via a scaling law for $M$.

\section{SSE Performance Bound for LISs and Optimal System Parameters}
We now analyze the performance bound of the SSE for a large (infinite) value for $M$.
As $M$ increases, the SSE converges to ${\bar \mu}_n^{\rm{SSE}}$
which depends on the value of ${\bar \mu}_{I_{nk}}/M^2$, as seen from (\ref{eq.MSSE}).
Hence, in an LIS system equipped with an infinite number of antennas, it is important to obtain its limiting value, $\mathop {\lim }\limits_{M \to \infty } \bar \mu _{{I_{nk}}}/{M^2}$,
and this can provide the performance bound of the SSE, asymptotically.
In this section, we derive the asymptotic SSE performance bound for an infinite $M$ using a scaling law of ${\bar \mu}_n^{\rm{SSE}}$,
and propose an optimal pilot training length based on that asymptotic bound.
We consider the uplink frame structure shown in Fig. {\ref{fig.2}}, in which all devices simultaneously transmit their orthogonal pilot sequences before transmitting their own data signals like a 3GPP model in \cite{ref.LTE2017TR36211}.
Due to the pilot training overhead $t$, there is a fundamental tradeoff between the received SINR enhancement and loss of uplink channel uses for the data signal.
Therefore, it is imperative to optimize the pilot training length to achieve the maximum SSE
and define how the optimal pilot training length deterministically scales with the various parameters of LIS system.

In order to derive the performance bound of LIS system, we first determine the scaling law of ${\bar \mu}_{I_{nk}}/M^2$ according to $M$.
Then, we have the following result related to the performance bound.

{\bf{{Theorem 2.}}} As $M$ increases, ${\bar \mu}_n^{\rm{SSE}}$ asymptotically converges to its performance bound, ${\hat \mu}_n^{\rm{SSE}}$, as given by:
${\bar \mu}_n^{\rm{SSE}} - {\hat \mu}_n^{\rm{SSE}} \xrightarrow[M \to \infty ]{} 0, $
where
\begin{align}
{\hat{\mu}_n^{{\rm{SSE}}}}
&=\left( {1 - \frac{t}{T}} \right)\sum\nolimits_{k = 1}^{K} {\log } \left( {1 + \frac{{{M^2\rho _{nk}}p_{nk}^2}}{{{\hat{\mu}}_{I_{nk}}}}} \right), \\
{{\hat{\mu}}_{I_{nk}}} &= {{{\rho _{nk}}{\left| {{\mu _{{x_{nk}}}}}\right|^2} + \sum\nolimits_{j \ne k}^{K} {{\rho _{nj}}{\left| {{\mu _{{y_{njk}}}}}\right|^2} }}  + \sum\nolimits_{l \ne n}^N {\sum\nolimits_{j = 1}^{K} {{\rho _{lj}}{\left| {{\mu _{{y_{lnjk}}}}}\right|^2} } } }. \label{eq.Mhat_Ink}
\end{align}
\begin{proof}
From (\ref{eq.MInk}), ${\bar \mu}_{I_{nk}}/M^2$ is obtained as follows:
\begin{equation}
\frac{{\bar\mu} _{{I_{nk}}}}{M^2}= {{{\rho _{nk}}\frac{{{\mu}}_{X_{nk}}}{M^2} + \sum\nolimits_{j \ne k}^{K} {{\rho _{nj}}\frac{{\bar{\mu}}_{Y_{njk}}}{M^2} }}  + \sum\nolimits_{l \ne n}^N {\sum\nolimits_{j = 1}^{K} {{\rho _{lj}}\frac{{\bar{\mu}}_{Y_{lnjk}}}{M^2}  + \frac{{{\mu}}_{Z_{nk}}}{M^2} } } }. \label{eq.MInk_M2}
\end{equation}
On the basis of Lemmas 1--3, we determine the scaling laws of the terms ${{{\mu}}_{X_{nk}}}/{M^2}$, ${{\bar{\mu}}_{Y_{njk}}}/{M^2}$, ${{\bar{\mu}}_{Y_{lnjk}}}/{M^2}$, and ${{{\mu}}_{Z_{nk}}}/{M^2}$ in (\ref{eq.MInk_M2}) according to $M$. 
As proved in Appendix D, the terms $\sigma _{{x_{nk}}}^2/M^2$, $\sigma _{{y_{njk}}}^2/M^2$, $\sigma _{{y_{lnjk}}}^2/M^2$, and $\sum\nolimits_{m=1}^M \sigma _{{z_{nkm}}}^2/M^2$ converge to zero as $M$ increases.
Similarly, the terms $\left| {{\mu _{{x_{nk}}}}}\right|^2/M^2$, $\left| {{\mu _{{y_{njk}}}}}\right|^2/M^2$, and $\left| {{\mu _{{y_{lnjk}}}}}\right|^2/M^2$ follow $\mathcal{O}({1})$, and $\sum\nolimits_{m=1}^M{\left| {{\mu _{{z_{nkm}}}}}\right|^2}/M^2$ decreases with $\mathcal{O}({1/M})$ as $M$ increases, based on the scaling laws for large $M$.
Therefore, we have the following asymptotic convergence:
\begin{equation}
\frac{{\mu} _{{X_{nk}}}}{M^2} - \frac{\left| {{\mu _{{x_{nk}}}}}\right|^2}{M^2} \xrightarrow[M \to \infty ]{} 0,\\
\end{equation}
\begin{equation}
\frac{{\bar\mu} _{{Y_{njk}}}}{M^2} - \frac{\left| {{\mu _{{y_{njk}}}}}\right|^2}{M^2} \xrightarrow[M \to \infty ]{} 0,\\
\end{equation}
\begin{equation}
\frac{{\bar\mu} _{{Y_{lnjk}}}}{M^2} - \frac{\left| {{\mu _{{y_{lnjk}}}}}\right|^2}{M^2} \xrightarrow[M \to \infty ]{} 0,\\
\end{equation}
\begin{equation}
\frac{{\mu} _{{Z_{nk}}}}{M^2}  \xrightarrow[M \to \infty ]{} 0,
\end{equation}
which completes the proof.
\end{proof}
The terms ${{\mu _{{x_{nk}}}}}$, ${{\mu _{{y_{njk}}}}}$, and ${{\mu _{{y_{lnjk}}}}}$ in (\ref{eq.Mhat_Ink}) are determined by the LOS channels depending on the locations of the devices, as shown, respectively, in (\ref{eq.A.Msxnk}), (\ref{eq.B.Msynjk}), and (\ref{eq.B.Msylnjk}). 
Therefore, the asymptotic SSE performance bound can be obtained, deterministically,
and that deterministic bound leads to \emph{several important implications when evaluating an LIS system, that significantly differ from conventional massive MIMO.}
First, an LIS system has a particular operating characteristic whereby the pilot contamination and intra/inter-LIS interference through the NLOS path and noise become negligible as $M$ increases.
If all of the inter-LIS interference is generated from the NLOS path, the pilot contamination and inter-LIS interference will vanish lead to a performance convergence between the SSE of single- and multi-LIS system. 
Moreover, unlike conventional massive MIMO in which the performance is dominantly limited by pilot contamination,
the channel estimation error including pilot contamination gradually loses its effect on the SSE,
and eventually, the SSE of a multi-LIS system will reach that of a single-LIS system with perfect CSI. 
More practically, even if all of the intra/inter-LIS interference channels are generated by device-specific spatially correlated Rician fading,
an LIS system also has a particular characteristic whereby its SSE performance is bounded by three factors that include pilot contamination, intra-, and inter-LIS interference through the LOS path.
\begin{table}[]
\centering\caption{Differences between massive MIMO and LIS systems}\vspace{-0.5cm}
\begin{tabular}{|c|l|l|}
\hline
\multicolumn{2}{|c|}{\textbf{System}}                                                        & \multicolumn{1}{c|}{\textbf{Description}}                                                                                                                                                                               \\ \hline
\multicolumn{2}{|c|}{Massive MIMO}                                                           & \begin{tabular}[c]{@{}l@{}}- Inter/intra-cell interference and noise are negligible.  \\ - SSE is limited by pilot contamination \cite{ref.Hoydis2013how}.\end{tabular}                                                                    \\ \hline
\multirow{5}{*}{LIS} & \begin{tabular}[c]{@{}l@{}}Rician ${\boldsymbol{h}}_{lnjk}$\\ \& Rician ${\boldsymbol{h}}_{nnjk}$\end{tabular}     & \begin{tabular}[c]{@{}l@{}}- Inter/intra-LIS interference through NLOS path and noise are negligible.\\ - SSE is limited by intra/inter-LIS interference and pilot contamination through LOS path.\end{tabular} \\ \cline{2-3} 
                     & \begin{tabular}[c]{@{}l@{}}Rayleigh ${\boldsymbol{h}}_{lnjk}$\\ \& Rician ${\boldsymbol{h}}_{nnjk}$\end{tabular}   & \begin{tabular}[c]{@{}l@{}}- Intra-LIS interference through NLOS path, inter-LIS interference, and noise are negligible.\\ - SSE is limited by intra-LIS interference through LOS path.\end{tabular}            \\ \cline{2-3} 
                     & \begin{tabular}[c]{@{}l@{}}Rayleigh ${\boldsymbol{h}}_{lnjk}$\\ \& Rayleigh ${\boldsymbol{h}}_{nnjk}$\end{tabular} & \begin{tabular}[c]{@{}l@{}}- All inter/intra-LIS interference and noise are negligible.\\ - SSE increases without bound as $M$ increases.\end{tabular}                                                                  \\ \hline
\end{tabular}\label{table.diff}\vspace{-0.5cm}
\end{table}
The differences between massive MIMO and our LIS are summarized in Table \ref{table.diff}.
Note that we can analyze the impact on the performance of the devices' positions based on the characteristic of the LOS path.
We define the distance from the center of desired LIS unit $k$ to that of interfering LIS unit $j$ as $d_{kj}^{\rm{L}}$.
Then, the transmit power at device $k$ at LIS $n$ is obtained as $\rho_{nk} = 4 \pi z_{nk}^2 \gamma_{\rm{t}}$
and the interference power received at LIS unit $j$ is obtained as $P_{kj}^{\rm{I}} = \frac{z_{nk}^2 \gamma_{\rm{t}}}{(d_{kj}^{\rm{L}})^2 + z_{nk}^2}$.
Since the first derivative of $P_{kj}^{\rm{I}}$ with respect to $z_{nk}$ is always larger than zero,
$P_{kj}^{\rm{I}}$ is a monotonically increasing function with respect to $z_{nk}$.
Therefore, the interference power increases as $z_{nk}$ increases while the desired signal power keeps constant, resulting in a performance degradation.

Next, we formulate an optimization problem whose goal is to maximize the SSE with respect to the pilot training length $t$
by using the asymptotic SSE, ${\bar \mu}_n^{\rm{SSE}}$, from Theorem 1.
Since ${\bar \mu}_{I_{nk}}$ is a function of $t$, we have
\begin{equation}
\hspace{-8.7cm}\mathop {\max }\limits_t \left( {1 - \frac{t}{T}} \right)\sum\limits_{k = 1}^{K} {\log \left( {1 + \frac{M^2{{\rho _{nk}}p_{nk}^2}}{{{\bar \mu}_{I_{nk}}(t)}}} \right)}, \label{eq.P1}\\
\end{equation}\addtocounter{equation}{-1}\begin{subequations}
\begin{equation}
\hspace{-13.7cm}{\rm{s.t.}}\ K \le t\le T,\vspace{-0.48cm}
\end{equation}
\begin{align}
{{\bar\mu} _{{I_{nk}}}}(t)&= {{\rho _{nk}}\left(\sigma _{{x_{nk}}}^2(t) + \left| {{\mu _{{x_{nk}}}}}\right|^2\right) + \sum\nolimits_{j \ne k}^{K} {{\rho _{nj}}\left(\sigma _{{y_{njk}}}^2 (t)+ \left| {{\mu _{{y_{njk}}}}}\right|^2\right)}}\nonumber\\
&+ \sum\nolimits_{l \ne n}^N {\sum\nolimits_{j = 1}^{K} {{\rho _{lj}}\left(\sigma _{{y_{lnjk}}}^2(t) + \left| {{\mu _{{y_{lnjk}}}}}\right|^2\right)  + \sum\nolimits_{m=1}^M\left({\sigma _{{z_{nkm}}}^2(t) + \left| {{\mu _{{z_{nkm}}}}}\right|^2}\right) } }. \label{eq.MInkt}
\end{align}
\end{subequations}

In (\ref{eq.MInkt}), the terms, $\sigma _{{x_{nk}}}^2(t)$, $\sigma _{{y_{njk}}}^2 (t)$, $\sigma _{{y_{lnjk}}}^2(t)$, and $\sigma _{{z_{nkm}}}^2(t)$, are monotonically decreasing functions with respect to $t$ as observed from (\ref{eq.A.Sxnk}), (\ref{eq.B.Synjk}), (\ref{eq.B.Sylnjk}), and (\ref{eq.C.Sznkm}), respectively.
Thus, ${{\bar\mu} _{{I_{nk}}}}(t)$ is also a monotonically decreasing function and ${\log \left( {1 + \frac{M^2{{\rho _{nk}}p_{nk}^2}}{{{\bar \mu}_{I_{nk}}(t)}}} \right)}$ is thus a positive concave increasing function with respect to $t$.
From the operations that preserve the concavity of functions \cite{ref.Boyd2004convex},
the product of a positive decreasing function and a positive concave increasing function is concave.
Thus, (\ref{eq.P1}) is a convex optimization problem and we can obtain the globally optimal result, ${t_{{\rm{opt}}}}$, using a simple gradient method.
Moreover, we can observe from the objective function in $(\ref{eq.P1})$ that ${t_{{\rm{opt}}}}$ changes according to deterministic values such as the locations of the devices and correlation matrices.
From Theorem 2, ${{\bar\mu} _{{I_{nk}}}}(t)$ asymptotically converges to ${{\hat\mu} _{{I_{nk}}}}$ as $M$ increases, resulting in the received SINR independent with $t$.
Therefore, in the following Corollary 1, we obtain an asymptotic value of $t_{\rm{opt}}$, independent of the locations of the devices and correlation matrices, using the asymptotic bound of the SSE from Theorem 2.

{\bf{{Corollary 1.}}} As $M$ increases, the optimal pilot training length can be asymptotically obtained as follows:
${t_{{\rm{opt}}}}-{K}\xrightarrow[M \to \infty ]{}0. $
\begin{proof}
As proved in Theorem 2, the SSE asymptotically converges to its performance bound ${\hat{\mu}_n^{{\rm{SSE}}}}$ as $M$ increases, and ${{\hat{\mu}}_{I_{nk}}}$ is independent with $t$ as seen from (\ref{eq.Mhat_Ink}).
Then, ${\hat{\mu}_n^{{\rm{SSE}}}}$ is a monotonically decreasing function with respect to $t$,
and therefore, the optimal pilot training length asymptotically converges to $K$, which completes the proof.
\end{proof}

\begin{table}[]
\caption{Simulation parameters}\vspace{-0.4cm}
\label{table.para}
\centering
\begin{tabular}{|c|c|c|c|}
\hline
\textbf{Parameter} & \textbf{Value} & \textbf{Parameter} & \textbf{Value}  \\ \hline
 Carrier frequency & $3$ GHz & Length of LIS unit $(2L)$ & $0.5$ $\rm{m}$ \\ \hline
Target SNR for uplink pilot & $0$ dB & Rician factor $({\kappa}[${dB}$])$ \cite{ref.LTE2017TR25996} & $13 - 0.03{d[\rm{m}]}$  \\ \hline
Target SNR for uplink data & $3$ dB & LOS path loss model \cite{ref.Tse2005fundamentals}& $11 +20{\rm{log}}_{10}d[\rm{m}]$ \\ \hline
Coherence block length $(T)$ & $500$ symbols & NLOS path loss model \cite{ref.Hoydis2013how} &  $37{\rm{log}}_{10}d[\rm{m}]$ $(\beta_{\rm{PL}}=3.7)$ \\ \hline
\end{tabular}\vspace{-0.2cm}
\end{table}

In conventional massive MIMO system, the pilot training length affects the received SINR and its optimal value is determined by various system parameters such as the number of users, uplink transmission period $T$, and transmit SNR, as proved in \cite{ref.Kim2018scaling}.
\emph{Unlike conventional massive MIMO},
Corollary 1 shows that, for both the single- and multi-LIS cases, 
the asymptotic bound of the received SINR does not increase with the increase in the pilot training length
and the maximum SSE can be achieved through a minimum pilot training length such as $t = K$ (i.e., one pilot symbol per device),
despite the pilot contamination effect.
With the proposed pilot training length $t = K$,
we then formulate an optimization problem that asymptotically maximizes the sum of SSE for neighboring $N$ LISs with respect to $K$,
as follows:
\begin{equation}
\mathop {\max }\limits_{K} \left( {1 - \frac{K}{T}} \right)\sum\limits_{n = 1}^{N} \sum\limits_{k = 1}^{K} {\log }\left(1+ {\hat{\gamma}}_{nk}(K)\right), \label{eq.P2}
\end{equation}\addtocounter{equation}{-1}\begin{subequations}
\begin{equation}
\hspace{-3cm}{\rm{s.t.}}\ 1 \le K \le T,
\end{equation}
\begin{equation}
\hspace{-1cm}{\hat{\gamma}}_{nk}(K) =  { \frac{{{M^2\rho _{nk}}p_{nk}^2}}{{{\hat{\mu}}_{I_{nk}}(K)}}}.
\end{equation}\end{subequations}

Consider that each device has its own prioritization coefficients for scheduling and each LIS schedules devices in order of their priority.
According to this priority order, the objective function in $(\ref{eq.P2})$ can be calculated from $K=1$ to $K=T$, deterministically,
given that the value of ${\hat{\gamma}}_{{nk}}(K)$ depends on the locations of the devices.
Therefore, by comparing those values over the entire ranges of $K$,
the optimal number of scheduled devices can be asymptotically derived, 
without the need for extensive simulations.
Note that an LIS has the potential for estimating the locations of serving devices from their uplink signal \cite{ref.Hu2018positioning}
and the Rician factor is calculated according to the distance between the LIS and device \cite{ref.LTE2017TR25996}.
By cooperating across adjacent multi-LISs like a network LIS, an LIS is able to share those information without the heavy burden of backhaul load and perform joint scheduling to maximize the network SE (NSE).
Therefore, the asymptotically optimal number of scheduled devices, $K_{\rm{opt}}$, can be calculated at each LIS, practically,
when adjacent multi-LISs cooperate with each other as a network LIS
and share the information about the locations of their own serving devices.

\section{Simulation Results and Analysis}
We present Monte Carlo simulation results for the uplink SSE in an LIS system,
and compare them with the results of the asymptotic analyses.
All simulations are statistically averaged over a large number of independent runs.
The simulation parameters are based on the LTE specifications, presented in Table \ref{table.para}.
In accordance with the LTE specifications \cite{ref.LTE2018TR36331}, the target SNR for uplink power control is semi-statically configured by upper-layer signaling in the LTE system. 
The range of target SNRs for the uplink data and pilot signals are -8 dB to 7 dB and -8 dB to 23 dB, respectively \cite{ref.LTE2018TR36331}.
The uplink target SNRs presented in Table \ref{table.para} satisfy the constraints of practical target SNRs.
Furthermore, the minimum scheduling unit is defined as 1 ms in the time domain, and 180 kHz in the frequency domain, which is the so-called physical resource block (PRB) in the LTE specifications \cite{ref.LTE2017TR36211}.
Each PRB consists of a total of 168 symbols (14 orthogonal frequency-division multiplexing (OFDM) symbols in the time domain, and 12 subcarriers in the frequency domain) including the cyclic prefix overhead of the OFDM symbols.
Since the value of $T$ = 500 presented in Table \ref{table.para} corresponds to approximately 3 PRBs ($500 \approx 168 \times 3$), 
we consider the uplink frame as one of two frames such as 1 ms $\times$ 540 kHz or 3 ms $\times$ 180 kHz.
The value of $T$ = 500 used in performance evaluation may be regarded as a moderate
coherence block length, given that the generalized coherence lengths are $T$ = 200 for high-mobility or high-delay spread
scenarios, and $T$ = 5000 for low-mobility or low-delay spread scenarios \cite{ref.Kim2018scaling}.

\begin{figure}[]
\centering
\includegraphics[width=0.8\columnwidth] {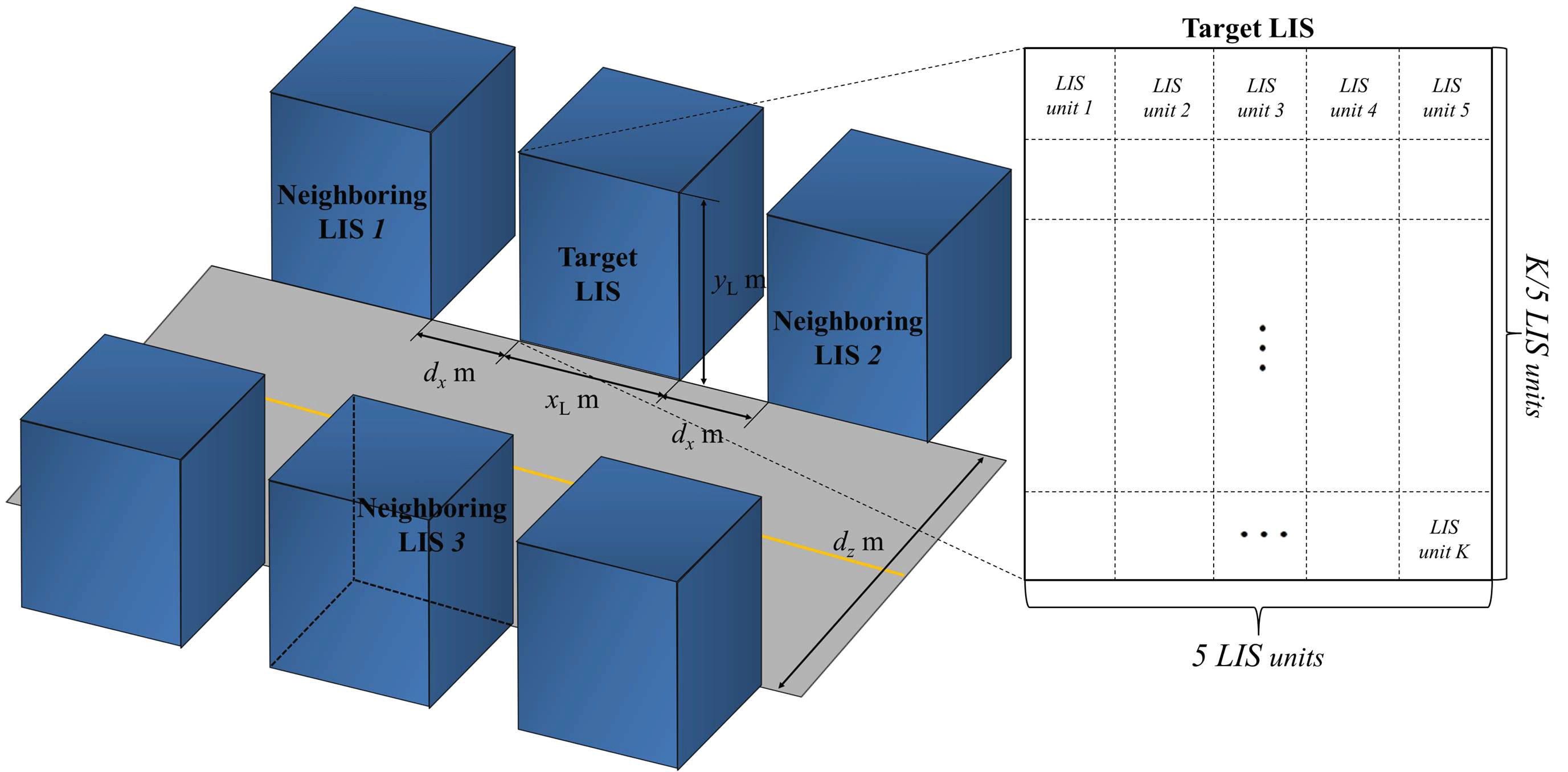}\vspace{-0.4cm}
\caption{Illustrative system model of the multiple LISs considered when $N=4$. }\vspace{-0.5cm}
\label{fig.3}
\end{figure}

In the simulations, we consider both single- and multi-LIS cases.
In both cases, we consider a scenario in which the devices are randomly and uniformly distributed within a three-dimensional space of $x_{\rm{L}}$ m$\times$ $y_{\rm{L}}$ m $\times$ $2$ m in front of each LIS.
For the uniform device distribution, we consider a simple grid approach that randomly selects $K$ points among the grid points in a three-dimensional space evenly divided by $0.1$ m intervals.
For the single-LIS case, $N=1$ and only a single target LIS is located in two-dimensional Cartesian space along the $xy$-plane.
For the multi-LIS case, to be able to consider the effect of pilot contamination, we assume a total of $N=4$ LISs consist of one target LIS and three neighboring LISs, located on both sides and in front of the target LIS, as shown in Fig. \ref{fig.3}.
Each LIS is divided into $K$ LIS units with $5\times K/5$ grid
and the parameters presented in Fig. \ref{fig.3} are such that $x_{\rm{L}} =10L$, $y_{\rm{L}} = 2LK/5$, $d_{x} = 4$, and $d_{z} = 6$.
Each device is assigned to each LIS unit depending on the distance between the center of each LIS unit and device.
All LISs are assumed to share the same frequency band, each of which serves $K$ devices and reuses $K$ pilot sequences.
In our simulation, all desired and interference channels are generated, respectively, based on the deterministic LOS and spatially correlated Rician fading.
In particular, each desired channel link is generated based on the LOS channel model given in (\ref{eq.1}) with the LOS path loss provided in Table \ref{table.para}.
Also, all interference channels (both inter- and intra-LIS interference) stem from device-specific spatially correlated Rician fading channels as given by (\ref{eq.2}) with the LOS and NLOS path loss provided in Table \ref{table.para}.
According to the 3GPP model in \cite{ref.LTE2017TR25996}, the existence of a LOS path depends on the distance from the transmitter and receiver.
The probability of a LOS is then given by
\begin{equation}
P_{lnjk}^{{\rm{LOS}}} = \left\{ \begin{array}{l}
\left( {{d_{\rm{C}}} - {d_{lnjk}}} \right)/{d_{\rm{C}}},{\rm{ }}0 < {d_{lnjk}} < {d_{\rm{C}}},\\
\,{\qquad\quad}0,\quad\qquad{d_{ljk}} > {d_C},
\end{array} \right.
\end{equation}
where ${d_{lnjk}}$ is the distance in meters between device $j$ at LIS $l$ and the center of the LIS unit $k$ part of LIS $n$, and ${d_{\rm{C}}}$ denotes the cutoff point, which is assumed to be $10$ m, as in \cite{ref.Jung2018lisul}. 
The Rician factor, $\kappa_{lnjk}$, is calculated according to ${d_{lnjk}}$, as per Table \ref{table.para}.

\begin{figure*}[t]\vspace{-2mm}
	\begin{multicols}{2}\vspace{-9mm}
		\hspace{-4.5mm}
		\includegraphics[width=1.1\columnwidth]{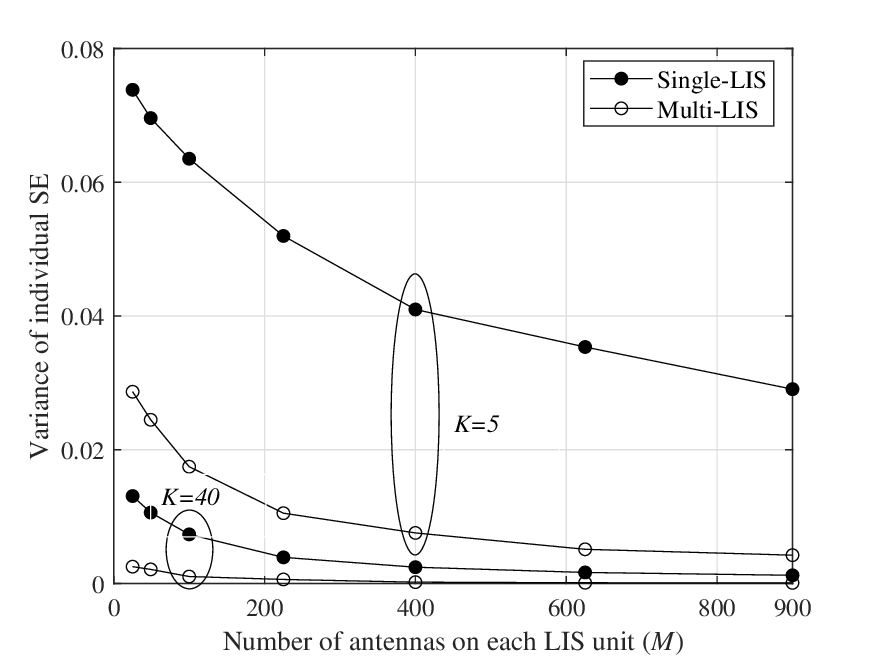}\vspace{-2.5mm}
		\par\caption{\small Variance of individual SE of an LIS system as a function of the number of antennas on the LIS unit.}
\label{fig.4}
		\hspace{-4.5mm}
		\includegraphics[width=1.1\columnwidth]{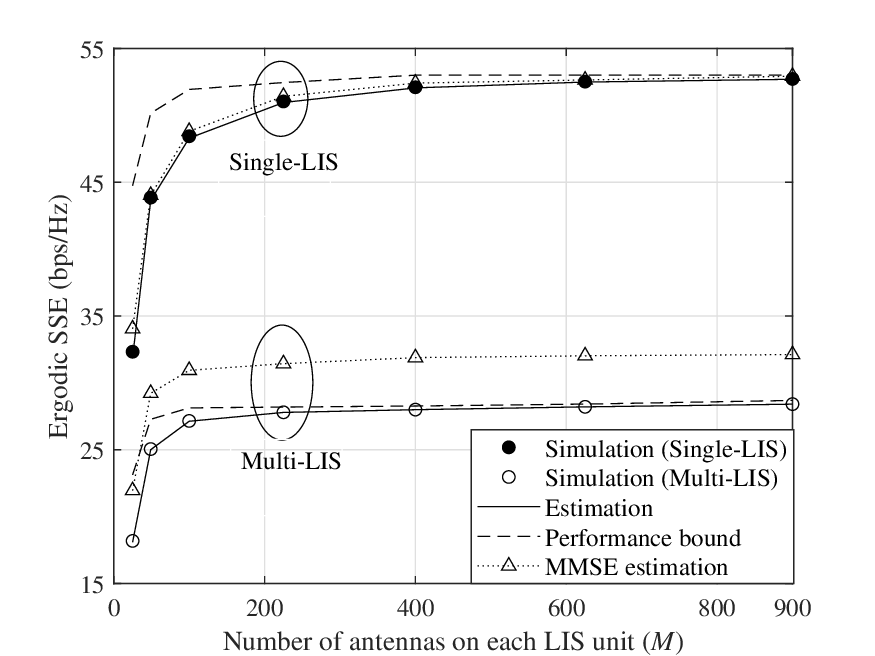}\vspace{-2.5mm}	
		\par\caption{\small Uplink ergodic SSE of an LIS system with Rician fading interference when $K=20$.}
\label{fig.5}
		\end{multicols}\vspace{-6mm}
\end{figure*}

Fig. \ref{fig.4} shows the channel hardening effect of an LIS system whereby the variances of individual SE in both single- and multi-LIS cases converge to zero as $M$ increases, despite the pilot contamination effect.
Here, the individual SE of device $K$ at LIS $n$ is derived from (\ref{eq.RSSE}) as follows:
\begin{equation}
{R_{nk}^{{\rm{SE}}}} =\left( {1 - \frac{t}{T}} \right){\log \left( {1 + \frac{{{\rho _{nk}}{S_{nk}}}}{{{I_{nk}}}}} \right). }
\end{equation}
The variance convergence of ${R_{nk}^{{\rm{SE}}}}$ verifies the asymptotic convergence of ${S_{nk}}/{I_{nk}}$,
and Lemma 4 is then verified given that $S_{nk}$ converges to a deterministic value as $M$ increases.

\begin{figure*}[t]\vspace{-2mm}
	\begin{multicols}{2}\vspace{-9mm}
		\hspace{-4.5mm}
		\includegraphics[width=1.1\columnwidth]{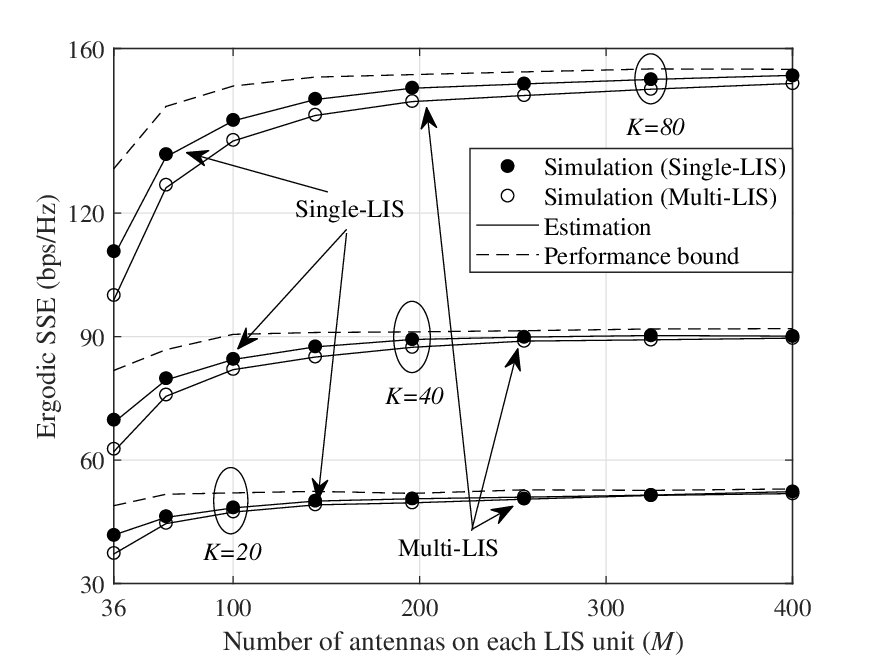}\vspace{-2.5mm}
		\par\caption{\small Uplink ergodic SSE of an LIS system with NLOS inter-LIS interference as a function of the number of antennas on the LIS unit.}
\label{fig.6}
		\hspace{-4.5mm}
		\includegraphics[width=1.1\columnwidth]{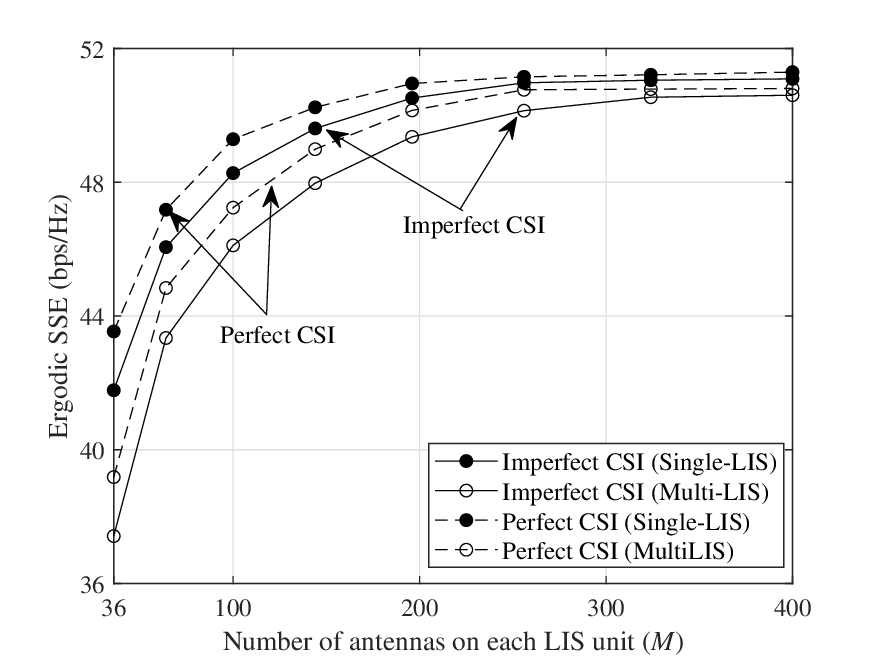}\vspace{-2.5mm}	
		\par\caption{\small Performance comparison of the ergodic SSE resulting from scenarios with perfect CSI and imperfect CSI when $K=20$ with NLOS inter-LIS interference.}
\label{fig.6_2}
		\end{multicols}\vspace{-6mm}
\end{figure*}

In Figs. \ref{fig.5}--\ref{fig.6_2}, Theorems 1 and 2 are verified in the following scenario.
All intra-LIS interference channels are generated by device-specific spatially correlated Rician fading.
In Fig. \ref{fig.5}, all inter-LIS interference channels are also generated by that Rician fading,
however, in Figs. \ref{fig.6} and \ref{fig.6_2}, those channels are generated entirely from the NLOS path such as spatially correlated Rayleigh fading.
In both Figs. \ref{fig.5} and \ref{fig.6}, the asymptotic results from Theorem 1 become close to the results of our simulations
and these results gradually approach to their performance bounds obtained from Theorem 2, as $M$ increases.
Moreover, those performance bounds also converge to the limiting values resulting from the intra/inter-LIS interference through the LOS path.
In Fig. \ref{fig.5}, the performance gap between the results of the single- and multi-LIS is roughly $21.3$ bps/Hz at $M=100$,
and it is expected to converge to $24.3$ bps/Hz from the bound gap between the two systems, as $M$ increases.
This performance gap between the two systems results from pilot contamination and inter-LIS interference generated from the LOS path, as proved in Theorem 2.
In Fig. \ref{fig.6}, the performance gap between the results of the single- and multi-LIS increases as $K$ increases from $K=20$ to $K=80$ because of the increase in the inter-LIS interference.
However, the performance gap between the two systems converges to zero even at $K=80$, as $M$ increases,
and their bounds achieve an equal performance over the entire range of $M$.
Since the the pilot contamination and inter-LIS interference generated from the NLOS path become negligible compared to the intra-LIS interference through the LOS path, this results in the performance convergence between the two systems
and eventually the multi-LIS system becomes an inter-LIS interference-free environment.

\begin{figure*}[t]\vspace{-2mm}
	\begin{multicols}{2}\vspace{-9mm}
		\hspace{-4.5mm}
		\includegraphics[width=1.1\columnwidth]{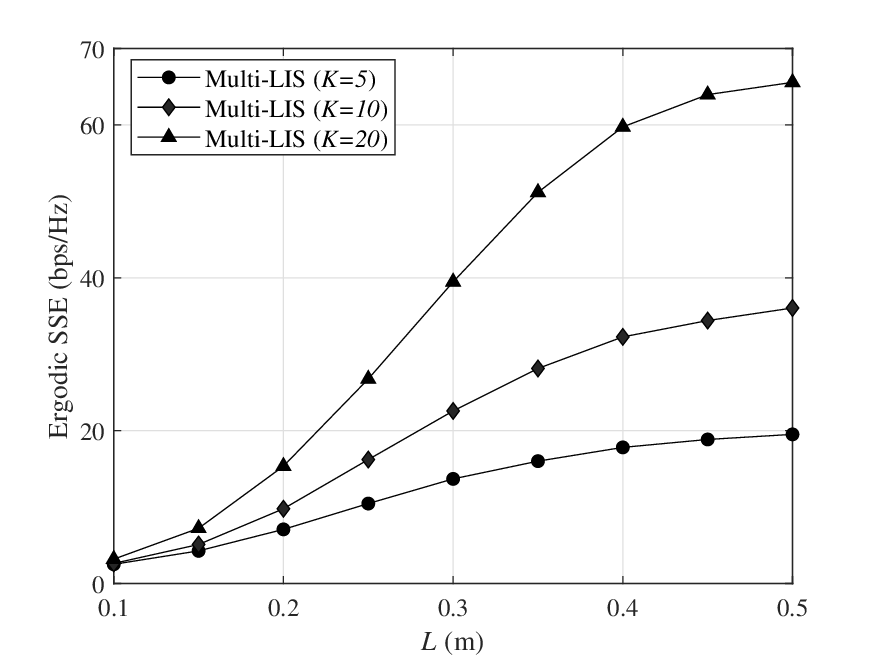}\vspace{-2.5mm}
		\par\caption{\small Uplink ergodic SSE of an LIS system as a function of $L$ when $M=100$.}
\label{fig.R1}
		\hspace{-4.5mm}
		\includegraphics[width=1.1\columnwidth]{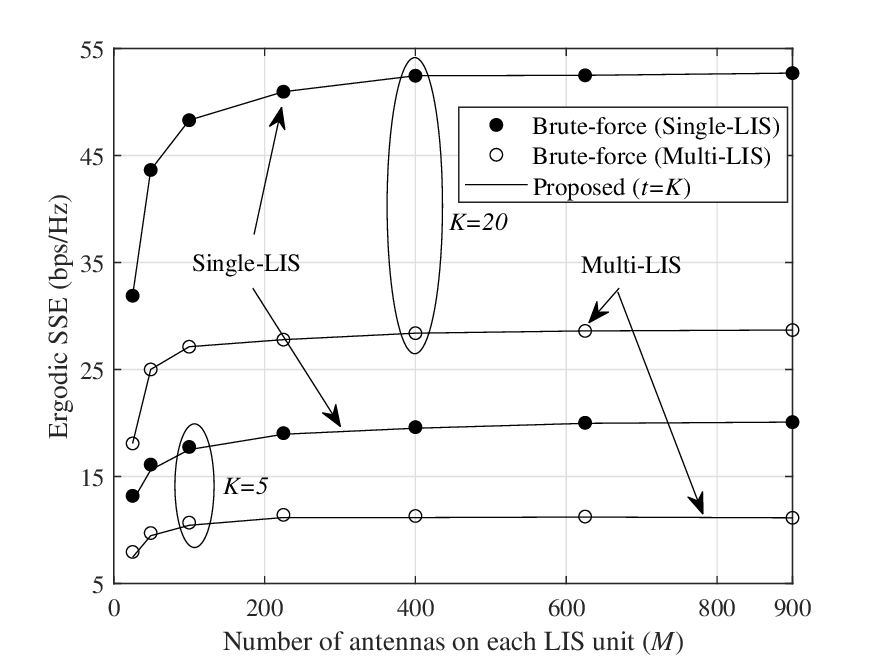}\vspace{-2.5mm}	
		\par\caption{\small Performance comparison between ergodic SSE resulting from the proposed $t$ and brute-force search.}
\label{fig.7}
		\end{multicols}\vspace{-8mm}
\end{figure*}

In case of multi-LIS, the SSE can be affected by both intra-LIS and inter-LIS interference, simultaneously.
However, in case of single-LIS, the SSE is only affected by intra-LIS interference.
Therefore, because of this interference factor, Fig. \ref{fig.5} shows that the ergodic SSE of the single-LIS case is higher than that of multi-LIS case.
Also, Fig. 5 shows the performance of the MMSE estimator as an upper bound of the considered LS estimator.
Since the MMSE estimator can reduce the pilot contamination through the LOS path compared to the LS estimator, 
in case of multi-LIS, the channel estimation error of the MMSE estimator is significantly reduced and the performance of the MMSE estimator is higher than that of the LS estimator.
However, the MMSE estimation is impractical for LIS systems because of the high processing complexity needed for estimating and sharing every correlation matrix.

Fig. \ref{fig.6_2} compares the ergodic SSE resulting from cases with perfect CSI and imperfect CSI, when $K=20$ and all inter-LIS interference channels are generated by spatially correlated Rayleigh fading.
We can observe that all ergodic SSE converge to same value of roughly $51$ bps/Hz.
Hence, despite the pilot contamination in the multi-LIS case, the ergodic SSE of the multi-LIS system with the imperfect CSI converges to that with the perfect CSI, and it eventually reaches the single-LIS performance with perfect CSI, as $M$ increases.
This clearly shows a particular characteristic of LIS systems whereby pilot contamination and inter-LIS interference become negligible,
representing a significant difference from conventional massive MIMO.

\begin{figure*}[t]\vspace{-2mm}
	\begin{multicols}{2}\vspace{-9mm}
		\hspace{-4.5mm}
		\includegraphics[width=1.1\columnwidth]{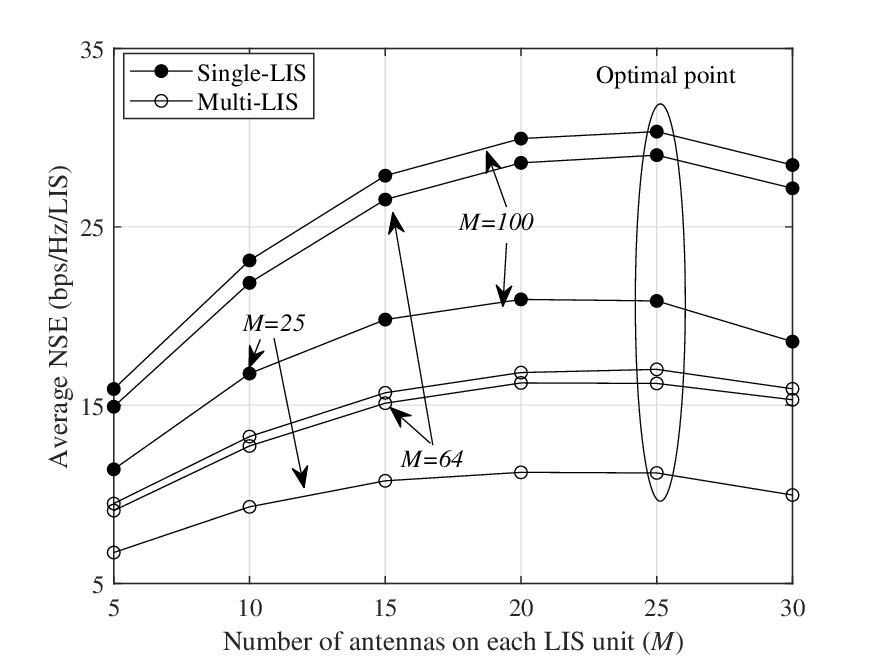}\vspace{-2.5mm}
		\par\caption{\small Average NSE of an LIS system as a function of the number of scheduled devices in each LIS when $T=50$ and $t=K$. }
\label{fig.8}
		\hspace{-4.5mm}
		\includegraphics[width=1.1\columnwidth]{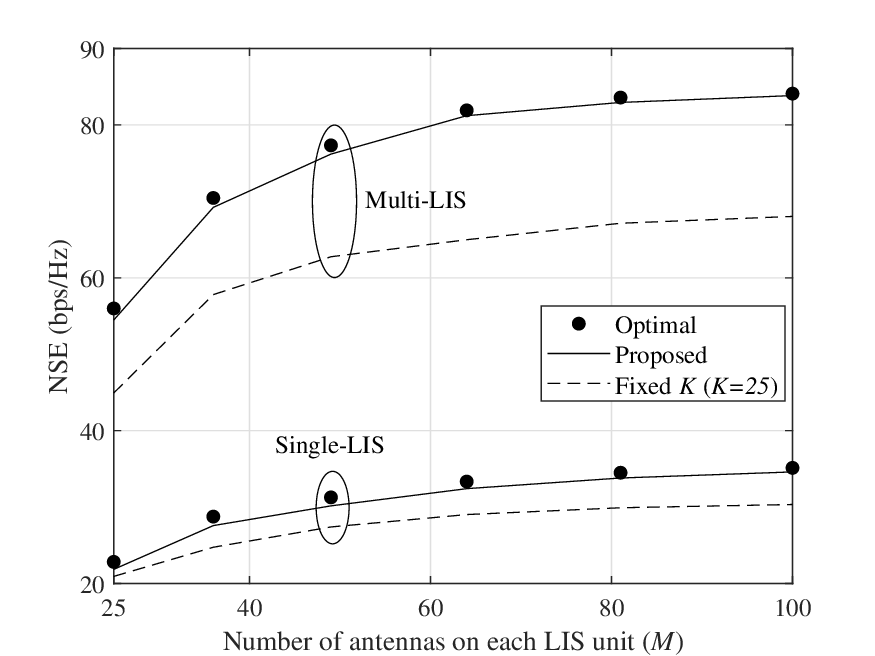}\vspace{-2.5mm}	
		\par\caption{\small Performance comparison between NSE resulting from the optimal $K$, proposed $K_{\rm{opt}}$, and fixed $K$, when $T=50$ and $t=K$.}
\label{fig.9}
		\end{multicols}\vspace{-8mm}
\end{figure*}

Fig. \ref{fig.R1} compares the ergodic SSE for different $K$ as a function of $L$ when $M=100$.
In case of multiple small LIS units, the SSE can be affected by both intra-LIS and inter-LIS interference, simultaneously.
However, in case of a single large LIS unit, the SSE is only affected by inter-LIS interference.
Note that an LIS is effective in intra-LIS interference suppression as proved in \cite{ref.Hu2018data} and \cite{ref.Hu2018assignment},
and the SSE is defined as the sum of $K$ individual SE of each device located within the target LIS.
Therefore, the SSE in case of multiple small LIS units is obtained by the sum of multiple SE
and will be higher than that of a single large LIS unit case, which is equal to the individual SE.
However, as $K$ increases given a limited area of practical LIS, 
inter-LIS interference will increase and the area of each LIS unit will decrease resulting in the desired signal power degradation, and this eventually results in lower individual SE.
Hence, there will be a fundamental tradeoff between the SSE and the individual SE in terms of $K$.
As shown in Fig. \ref{fig.R1}, the SSE in case of $K=20$ is always much higher than that of $K=5$, regardless of $L$.
However, when $L = 0.5$ m, the average individual SE in case of $K=20$ is $3.28$ bps/Hz which is lower than that of $K=5$ which is $3.9$ bps/Hz.
Moreover, since inter-LIS interference exclusively depends on $K$ regardless of $L$,
the SSE increases as $L$ increases as shown in Fig. \ref{fig.R1}.

Fig. \ref{fig.7} compares the ergodic SSE resulting from the optimal training lengths obtained from Corollary 1 and a brute-force search, as a function of $M$.
As shown in Fig. \ref{fig.7}, the optimal performance obtained by a brute-force search is nearly achieved by the proposed pilot training length, $t=K$, over the entire range of $M$.
Although the ergodic SSE of a multi-LIS system is affected by pilot contamination and inter-LIS interference through the LOS path, 
thus resulting in performance degradation compared with the single-LIS case,
the minimum training length always achieves the optimal performance, regardless of the number of neighboring LISs and devices located within their serving area.
This result shows a particular characteristic of an LIS
that the accurate CSI is not an important system parameter in both single- and multi-LIS cases, unlike conventional massive MIMO.

Fig. \ref{fig.8} shows the average NSE with the proposed pilot training length as a function of $K$, 
when $T=50$, considering very high-mobility scenarios.
The average NSE is defined by the sum of ergodic SSE for $N$ LISs divided by $N$.
Since the pilot training length $t$ does not affect the asymptotic received SINR as proved in Corollary 1,
the uplink channel uses for the data signal and its SINR decrease as $K$ increases 
due to the the increase in the pilot training overhead and the intra/inter-LIS interference, respectively.
Meanwhile, the NSE improves as $K$ increases given that it stems from the sum rate of $NK$ devices
located within the serving area of a network LIS.
Therefore, a fundamental tradeoff exists in terms of the average NSE according to the value of $K$.
Due to the logarithmic nature of the mutual information, the average NSE increases with $K$ when $K$ is small, and starts decreasing with $K$ when $K$ exceeds a given threshold point, as shown in Fig. \ref{fig.8}.
The maximum NSE can be achieved, statistically, by the optimal value of $K$,
which could be obtained experimentally as $K=25$.

Fig. \ref{fig.9} compares the total NSE with the proposed number of scheduled devices, $K_{\rm{opt}}$,
and that with a fixed number of scheduled devices, as a function of $M$.
This fixed number is obtained experimentally as $K=25$ from Fig. \ref{fig.8},
and $K_{\rm{opt}}$ can be obtained deterministically according to each device distribution.
The optimal performance is also presented in Fig. \ref{fig.9}. 
This optimal value is obtained, experimentally, by comparing every NSE over the entire ranges of $K$ for each device distribution.
Fig. \ref{fig.9} shows that the NSE with the proposed $K_{\rm{opt}}$ is always higher than that with $K=25$ over the entire ranges of $M$
and their performance gap increases as $M$ increases.
Moreover, the NSE with $K_{\rm{opt}}$ nearly achieves the optimal performance obtained from our extensive simulations.

\section{Conclusions}
In this paper, we have asymptotically analyzed the performance of an LIS system under practical LIS environments with a well-defined uplink frame structure and the pilot contamination. 
In particular, we have derived the asymptotic SSE by considering a practical LIS environment in which the interference channels are generated by device-specific spatially correlated Rician fading and channel estimation errors can be caused by pilot contamination based on a practical uplink frame structure.
We have shown that the asymptotic results can accurately and analytically determine the performance of an LIS without the need for extensive simulations.
Moreover, we have studied the performance bound of SSE from the derived asymptotic SSE and obtained the optimal pilot training length to maximize the SSE, showing
that the maximum SSE can be achieved with a minimum pilot training length of $t=K$, regardless of the pilot contamination effect.
Furthermore, we have proved that the SSE of a multi-LIS system is bounded by three factors: pilot contamination, intra-LIS interference, and inter-LIS interference generated from the LOS path.
On the other hand, the pilot contamination and intra/inter-LIS interference generated from the NLOS path and noise become negligible as $M$ increases. 
Simulation results have shown that our analytical results are in close agreement with the results arising from extensive simulations.
Our results also show that, unlike conventional massive MIMO system, the effect of pilot contamination has been shown to become negligible when the inter-LIS interference is generated from NLOS path.
Moreover, we have observed that the SSE of the proposed pilot training lengths achieve those obtained with the optimal lengths determined by a brute-force search, both in single- and multi-LIS environments.
Furthermore, the maximum value of the NSE has been shown to be achievable, practically, by using the proposed number of scheduled devices based on a network LIS.
In summary, in order to properly conduct the standardization process for LIS systems, it will be necessary to take into account the need for an adequate frame structure including the proposed pilot training length and the number of scheduled devices.

\section*{Appendix A\\  Proof of Lemma 1}
Given the definition of $X_{nk}= {{\left| {{\boldsymbol{e}}_{nk}^{\rm{H}}{{\boldsymbol{h}}_{nnkk}^{\rm{L}}}} \right|}^2}$,
we define
\begin{equation}
x_{nk}={{\boldsymbol{e}}_{nk}^{\rm{H}}{{\boldsymbol{h}}_{nnkk}^{\rm{L}}}}
= \sum\nolimits_{l \ne n}^{N} {\sqrt {\frac{{{\rho _{{{\rm{p}}_{lk}}}}}}{{{\rho _{{\rm{p}}_{nk}}}}}} \left( {{{{\boldsymbol{\bar h}}}_{lnkk}^{\rm{H}}}{{\boldsymbol{h}}_{nnkk}^{\rm{L}}} + {{{\boldsymbol{\tilde h}}}_{lnkk}^{\rm{H}}}{\boldsymbol{h}}_{nnkk}^{\rm{L}}} \right)}  + \frac{1}{{\sqrt {t{\rho _{{\rm{p}}_{nk}}}} }}{{\boldsymbol{w}}_{nk}^{\rm{H}}}{{\boldsymbol{h}}_{nnkk}^{\rm{L}}}. 
\end{equation}
Since ${{{\boldsymbol{\bar h}}}_{lnkk}^{\rm{H}}}{{\boldsymbol{h}}_{nnkk}^{\rm{L}}}$ is deterministic value without any variance,
the terms ${{{\boldsymbol{\tilde h}}}_{lnkk}^{\rm{H}}}{{\boldsymbol{h}}_{nnkk}^{\rm{L}}}$ and ${{\boldsymbol{w}}_{nk}^{\rm{H}}}{{\boldsymbol{h}}_{nnkk}^{\rm{L}}}$ determine the distribution of $x_{nk}$.
From \cite{ref.Jung2018lisul}, the terms ${{{\boldsymbol{\tilde h}}}_{lnkk}^{\rm{H}}}{{\boldsymbol{h}}_{nnkk}^{\rm{L}}}$ and ${{\boldsymbol{w}}_{nk}^{\rm{H}}}{{\boldsymbol{h}}_{nnkk}^{\rm{L}}}$ follow a complex Gaussian distribution as follows:
\begin{align}
{{{\boldsymbol{\tilde h}}}_{lnkk}^{\rm{H}}}{{\boldsymbol{h}}_{nnkk}^{\rm{L}}} &\sim \mathcal{CN}\left( {{0}},\frac{1}{{{{\kappa _{lnkk}} + 1}}}\sum\nolimits_{p = 1}^P {\left| {{\boldsymbol{c}}_{lnkkp}^{\rm{H}}{{\boldsymbol{h}}_{nnkk}^{\rm{L}}}} \right|}^2 \right),\label{eq.A.NL}\\
{{\boldsymbol{w}}_{nk}^{\rm{H}}}{{\boldsymbol{h}}_{nnkk}^{\rm{L}}} &\sim \mathcal{CN}\left( {{0}},\sum\nolimits_{m = 1}^M \left({\beta _{nnkkm}^{\rm{L}}}\right)^2 \right).
\end{align}
Since ${{{\boldsymbol{\tilde h}}}_{lnkk}^{\rm{H}}}{{\boldsymbol{h}}_{nnkk}^{\rm{L}}}$ and ${{\boldsymbol{w}}_{nk}^{\rm{H}}}{{\boldsymbol{h}}_{nnkk}^{\rm{L}}}$ are independent random variables, $x_{nk}$ also follows a complex Gaussian distribution as follows:
$x_{nk} \sim \mathcal{CN}\left( {{\mu_{x_{nk}}}},{{\sigma_{x_{nk}}^{2}}}\right)$, where
\begin{equation}
{{\mu_{x_{nk}}}} = \sum\nolimits_{l \ne n}^{N} {\sqrt {\frac{{{\rho _{{{\rm{p}}_{lk}}}}}}{{{\rho _{{\rm{p}}_{nk}}}}}}  {{{{\boldsymbol{\bar h}}}_{lnkk}^{\rm{H}}}{{\boldsymbol{h}}_{nnkk}^{\rm{L}}} }},\label{eq.A.Msxnk}
\end{equation}
\begin{equation}
{{\sigma_{x_{nk}}^{2}}} = \sum\nolimits_{l \ne n}^{N} { {\frac{{{\rho _{{{\rm{p}}_{lk}}}}}}{{{\rho _{{\rm{p}}_{nk}}}\left({{{\kappa _{lnkk}} + 1}}\right)}}}   \sum\nolimits_{p = 1}^P {\left| {{\boldsymbol{c}}_{lnkkp}^{\rm{H}}{{\boldsymbol{h}}_{nnkk}^{\rm{L}}}} \right|}^2}   + \frac{1}{{ {t{\rho _{{\rm{p}}_{nk}}}} }}\sum\nolimits_{m = 1}^M \left({\beta _{nnkkm}^{\rm{L}}}\right)^2.\label{eq.A.Sxnk}
\end{equation}
From the definition of $X_{nk}= {{\left| x_{nk} \right|}^2}$, the mean of ${X_{nk}}$ can be obtained by 
${{\mu} _{{X_{nk}}}}=\sigma _{{x_{nk}}}^2 + \left| {{\mu _{{x_{nk}}}}}\right|^2$.

\section*{Appendix B\\ Proof of Lemma 2}
Given the definition of ${Y_{njk}} = {{\left| {{\boldsymbol{\hat h}}_{nnkk}^{\rm{H}}{{\boldsymbol{h}}_{nnjk}}} \right|}^2}$, we define
\begin{equation}
y_{njk} = {{\boldsymbol{\hat h}}_{nnkk}^{\rm{H}}{{\boldsymbol{h}}_{nnjk}}}
={({\boldsymbol{h}}_{nnkk}^{\rm{L}})^{\rm{H}}{{\boldsymbol{\bar h}}_{nnjk}}} + {{\boldsymbol{e}}_{nk}^{\rm{H}}{{\boldsymbol{\bar h}}_{nnjk}}} + 
{\left({\boldsymbol{e}}_{nk}^{\rm{H}}+({\boldsymbol{h}}_{nnkk}^{\rm{L}})^{\rm{H}}\right){{\boldsymbol{\tilde h}}_{nnjk}}}.\label{eq.B.synjk} 
\end{equation}
For notational simplicity, we define that $y_{njk}^{\rm{LL}} = {({\boldsymbol{h}}_{nnkk}^{\rm{L}})^{\rm{H}}{{\boldsymbol{\bar h}}_{nnjk}}}$,
$ y_{njk}^{\rm{EL}}={{\boldsymbol{e}}_{nk}^{\rm{H}}{{\boldsymbol{\bar h}}_{nnjk}}}$, and
$y_{njk}^{\rm{EN}} = {\left({\boldsymbol{e}}_{nk}^{\rm{H}}+({\boldsymbol{h}}_{nnkk}^{\rm{L}})^{\rm{H}}\right){{\boldsymbol{\tilde h}}_{nnjk}}}$.
Then, $y_{njk}^{\rm{LL}}$ is the deterministic value depending on the locations of the devices.
Also, $y_{njk}^{\rm{EL}}$ is obtained similarly as $x_{nk}$ as follows:
$y_{njk}^{\rm{EL}} \sim \mathcal{CN}\big( {\mu_{x_{njk}^{\rm{EL}}}},{{\sigma_{x_{njk}^{\rm{EL}}}^{2}}}\big),$
where
\begin{align}
{\mu_{y_{njk}^{\rm{EL}}}}&=\sum\nolimits_{l \ne n}^{N} {\sqrt {\frac{{{\rho _{{{\rm{p}}_{lk}}}}}}{{{\rho _{{\rm{p}}_{nk}}}}}}  {{{{\boldsymbol{\bar h}}}_{lnkk}^{\rm{H}}}}{{{{\boldsymbol{\bar h}}}_{nnjk}}} },\\
\sigma _{{y_{njk}^{\rm{EL}}}}^2  &= \sum\nolimits_{l \ne n}^{N} {\frac{{{\rho_{{\rm{p}}_{lk}}}}}{{{{\rho _{{{\rm{p}}_{nk}}}\left(\kappa _{lnkk}+1\right)}}}}} \sum\nolimits_{p = 1}^P {\left| {{\boldsymbol{c}}_{lnkkp}^{\rm{H}}{{{\boldsymbol{\bar h}}}_{nnjk}}} \right|}^2  + \frac{1}{{t{\rho _{{{\rm{p}}_{nk}}}}}}\sum\nolimits_{m = 1}^M \left({\beta _{nnjkm}^{\rm{L}}}\right)^2.\label{eq.B.ynjk_EL}
\end{align}
Next, a random variable $y_{njk}^{\rm{EN}}$ can be expressed as follows:
\begin{equation}
y_{njk}^{\rm{EN}}=  
{\boldsymbol{\bar q}}_{nk} ^{\rm{H}}{{{{\boldsymbol{ \tilde h}}}_{nnjk}}}
+
\sum\nolimits_{l \ne n}^{N} {\sqrt {\frac{{{\rho _{{{\rm{p}}_{lk}}}}}}{{{\rho _{{\rm{p}}_{nk}}}}}}   {{{{\boldsymbol{\tilde h}}}_{lnkk}^{\rm{H}}}}{{{{\boldsymbol{\tilde h}}}_{nnjk}}}} 
+ \frac{{{\boldsymbol{w}}_{nk}^{\rm{H}}}{{\boldsymbol{\tilde h}}_{nnjk}}}{{\sqrt {t{\rho _{{\rm{p}}_{nk}}}} }},\label{eq.B.yEN}
\end{equation}
where
${\boldsymbol{\bar q}}_{nk} = {\boldsymbol{h}}_{nnkk}^{\rm{L}} +\sum\nolimits_{l \ne n}^{N} {\sqrt {\frac{{{\rho _{{{\rm{p}}_{lk}}}}}}{{{\rho _{{\rm{p}}_{nk}}}}}}  {{{{\boldsymbol{\bar h}}}_{lnkk}}}}.$
The terms ${{{{\boldsymbol{\bar q}}}_{nk}^{\rm{H}}}}{{{{\boldsymbol{ \tilde h}}}_{nnjk}}}$ and 
${{\boldsymbol{w}}_{nk}^{\rm{H}}}{{\boldsymbol{\tilde h}}_{nnjk}}$ in (\ref{eq.B.yEN}) can be obtained, respectively, similarly as (\ref{eq.A.NL}) and using the Lyapunov central limit theorem from \cite{ref.Jung2018lisul}:
\begin{gather}
{{{{\boldsymbol{\bar q}}}_{nk}^{\rm{H}}}}{{{{\boldsymbol{ \tilde h}}}_{nnjk}}} \sim \mathcal{CN}\left( 0,{\frac{1}{{{{\kappa _{nnjk}+1}}}}} {\sum\nolimits_{p = 1}^P {\left| {{{{\boldsymbol{\bar q}}}_{nk}^{\rm{H}}}{\boldsymbol{c}}_{nnjkp}} \right|}^2
}\right),\\
\sqrt{\frac{{Mt{\rho _{{{\rm{p}}_{nk}}}}}({{\kappa _{nnjk}+1}})}{\sum\nolimits_{m,p}^{M,P} {{{\left( {\alpha _{nnjkp}^{{\rm{NL}}}l_{nnjkm}^{{\rm{NL}}}} \right)}^2}}}}{{\boldsymbol{w}}_{nk}^{\rm{H}}}{{\boldsymbol{\tilde h}}_{nnjk}} \xrightarrow[\hspace{-2pt}M \to \infty \hspace{-2pt}]{\rm{d}} \mathcal{CN}\left( 0,1 \right), \label{eq.B.wh}
\end{gather}
where `` $\xrightarrow[\hspace{-2pt}M \to \infty \hspace{-2pt}]{\rm{d}}$'' denotes the convergence in distribution.
Also, ${{{{\boldsymbol{\tilde h}}}_{lnkk}^{\rm{H}}}}{{{{\boldsymbol{\tilde h}}}_{nnjk}}}$ in (\ref{eq.B.yEN})
is given by
\begin{equation}
{{{{\boldsymbol{\tilde h}}}_{lnkk}^{\rm{H}}}}{{{{\boldsymbol{\tilde h}}}_{nnjk}}}
= \frac{{\boldsymbol{g}}_{lnkk}^{\rm{H}} \Big({{{\big( {{\boldsymbol{R}}_{lnkk}^{1/2}} \big)}^{\rm{H}}}{\boldsymbol{R}}_{nnjk}^{1/2}}\Big) {\boldsymbol{g}}_{nnjk}}{\sqrt{\left(\kappa _{lnkk}+1\right)\left(\kappa _{nnjk}+1\right)}}.
\end{equation}
Given random vectors ${\boldsymbol{g}}_{lnkk}$ and ${\boldsymbol{g}}_{nnjk}$,
those elements are independent each other and identically follow $\mathcal{CN}\left( {{0}},1 \right)$.
Similarly as (\ref{eq.B.wh}), we have
\begin{equation}
\frac{\sqrt{\left(\kappa _{lnkk}+1\right)\left(\kappa _{nnjk}+1\right)}}{\Big\| {{{\big( {{\boldsymbol{R}}_{lnkk}^{1/2}} \big)}^{\rm{H}}}{\boldsymbol{R}}_{nnjk}^{1/2}} \Big\|_{\rm{F}}}
{{{{\boldsymbol{\tilde h}}}_{lnkk}^{\rm{H}}}}{{{{\boldsymbol{\tilde h}}}_{nnjk}}}  \xrightarrow[\hspace{-2pt}M \to \infty \hspace{-2pt}]{\rm{d}}\mathcal{CN} \left(0,  1\right).
\end{equation}
Since the terms ${{{{\boldsymbol{\bar q}}}_{nk}^{\rm{H}}}}{{{{\boldsymbol{ \tilde h}}}_{nnjk}}}$, ${{{{\boldsymbol{\tilde h}}}_{lnkk}^{\rm{H}}}}{{{{\boldsymbol{\tilde h}}}_{nnjk}}}$, and ${{\boldsymbol{w}}_{nk}^{\rm{H}}}{{\boldsymbol{\tilde h}}_{nnjk}}$ in (\ref{eq.B.yEN}) are independent of each other, we have the following convergence in distribution:
$\frac{1}{{{\sigma_{y_{njk}^{\rm{EN}}}}}}y_{njk}^{\rm{EN}} \xrightarrow[\hspace{-2pt}M \to \infty \hspace{-2pt}]{\rm{d}} \mathcal{CN}\left( 0,1\right),$
where 
\begin{equation}
\sigma _{{y_{njk}^{\rm{EN}}}}^2 =  \frac{1}{\kappa _{nnjk}+1}\left( \sum\limits_{p = 1}^P {\left| {{{{\boldsymbol{\bar q}}}_{nk}^{\rm{H}}}{\boldsymbol{c}}_{nnjkp}} \right|}^2 
+\sum\limits_{l \ne n}^{N } {
\frac{{{{\rho_{{\rm{p}}_{lk}}}}}\Big\| {{{\big( {{\boldsymbol{R}}_{lnkk}^{1/2}} \big)}^{\rm{H}}}{\boldsymbol{R}}_{nnjk}^{1/2}} \Big\|_{\rm{F}}^2}{{{{\rho _{{{\rm{p}}_{nk}}}}}}(\kappa _{lnkk}+1)}}
 + \sum\limits_{m,p}^{M,P} \frac{{{\left( {\alpha _{nnjkp}^{{\rm{NL}}}l_{nnjkm}^{{\rm{NL}}}} \right)}^2}}{{Mt{\rho _{{{\rm{p}}_{nk}}}}}}\right)
.\label{eq.B.ynjk_EN}
\end{equation}
We can observe from (\ref{eq.B.synjk}) that $y_{njk}^{\rm{LL}}$, $ y_{njk}^{\rm{EL}}$, and $y_{njk}^{\rm{EN}}$ are independent of each other.
Thus, $y_{njk}$ asymptotically follows:
$\frac{1}{{{\sigma_{y_{njk}}}}}\big(y_{njk} - {\mu_{y_{njk}}}\big) \xrightarrow[\hspace{-2pt}M \to \infty \hspace{-2pt}]{\rm{d}} \mathcal{CN}\left( 0,1\right),$
where
\begin{align}
{{\mu _{{y_{njk}}}}} &= ({\boldsymbol{h}}_{nnkk}^{\rm{L}})^{\rm{H}}{\bar{\boldsymbol{h}}}_{nnjk} 
+ \sum\limits_{l \ne n}^{N} {\sqrt {\frac{{{\rho _{{{\rm{p}}_{lk}}}}}}{{{\rho _{{\rm{p}}_{nk}}}}}}  {{{{\boldsymbol{\bar h}}}_{lnkk}^{\rm{H}}}}{{{{\boldsymbol{\bar h}}}_{nnjk}}} }, \label{eq.B.Msynjk}\\
\sigma _{{y_{njk}}}^2  &= \sigma _{{y_{njk}^{\rm{EL}}}}^2 + \sigma _{{y_{njk}^{\rm{EN}}}}^2. \label{eq.B.Synjk}
\end{align}
From the definition of $Y_{njk}= {{\left| y_{njk} \right|}^2}$, the mean of ${Y_{njk}}$ follows
${{\mu} _{{Y_{njk}}}}- {{\bar\mu} _{{Y_{njk}}}}\xrightarrow[\hspace{-2pt}M \to \infty \hspace{-2pt}]{}0$, where
\begin{equation}
{{\bar\mu} _{{Y_{njk}}}}=\sigma _{{y_{njk}}}^2 + \left| {{\mu _{{y_{njk}}}}}\right|^2. \label{eq.B.MYnjk}
\end{equation}
Similarly, given that ${Y_{lnjk}} = {{\Big| {{\boldsymbol{\hat h}}_{nnkk}^{\rm{H}}{{\boldsymbol{h}}_{lnjk}}} \Big|}^2}$
and $y_{lnjk} = {{\boldsymbol{\hat h}}_{nnkk}^{\rm{H}}{{\boldsymbol{h}}_{lnjk}}}$, the mean of $Y_{lnjk}$ follows
${{\mu} _{{Y_{lnjk}}}}- {{\bar\mu} _{{Y_{lnjk}}}}\xrightarrow[\hspace{-2pt}M \to \infty \hspace{-2pt}]{}0$, where
\begin{equation}
{{\bar\mu} _{{Y_{lnjk}}}}=\sigma _{{y_{lnjk}}}^2 + \left| {{\mu _{{y_{lnjk}}}}}\right|^2, \label{eq.B.MYlnjk}\\
\end{equation}
\begin{align}
{{\mu _{{y_{lnjk}}}}} &= ({\boldsymbol{h}}_{nnkk}^{\rm{L}})^{\rm{H}}{\bar{\boldsymbol{h}}}_{lnjk} 
+ \sum\limits_{l \ne n}^{N} {\sqrt {\frac{{{\rho _{{{\rm{p}}_{lk}}}}}}{{{\rho _{{\rm{p}}_{nk}}}}}}  {{{{\boldsymbol{\bar h}}}_{lnkk}^{\rm{H}}}}{{{{\boldsymbol{\bar h}}}_{lnjk}}} }, \label{eq.B.Msylnjk}\\
\sigma _{{y_{lnjk}}}^2  &= \sigma _{{y_{lnjk}^{\rm{EL}}}}^2  + \sigma _{{y_{lnjk}^{\rm{EN}}}}^2,\label{eq.B.Sylnjk}\\
\sigma _{{y_{lnjk}^{\rm{EL}}}}^2 & = \sum\limits_{l \ne n}^{N} {\frac{{{\rho_{{\rm{p}}_{lk}}}}}{{{{\rho _{{{\rm{p}}_{nk}}}\left(\kappa _{lnkk}+1\right)}}}}} \sum\limits_{p = 1}^P {\left| {{\boldsymbol{c}}_{lnkkp}^{\rm{H}}{{{\boldsymbol{\bar h}}}_{lnjk}}} \right|}^2  + \frac{1}{{t{\rho _{{{\rm{p}}_{nk}}}}}}\sum\limits_{m = 1}^M \left({\beta _{lnjkm}^{\rm{L}}}\right)^2,
\end{align}
\begin{align}
\sigma _{{y_{lnjk}^{\rm{EN}}}}^2 &=  \frac{1}{\kappa _{lnjk}+1}\left( \sum\limits_{p = 1}^P {\left| {{{{\boldsymbol{\bar q}}}_{nk}^{\rm{H}}}{\boldsymbol{c}}_{lnjkp}} \right|}^2 
+\sum\limits_{l \ne n}^{N } {
\frac{{{{\rho_{{\rm{p}}_{lk}}}}}\Big\| {{{\big( {{\boldsymbol{R}}_{lnkk}^{1/2}} \big)}^{\rm{H}}}{\boldsymbol{R}}_{lnjk}^{1/2}} \Big\|_{\rm{F}}^2}{{{{\rho _{{{\rm{p}}_{nk}}}}}}(\kappa _{lnkk}+1)}}
 + \sum\limits_{m,p}^{M,P} \frac{{{\left( {\alpha _{lnjkp}^{{\rm{NL}}}l_{lnjkm}^{{\rm{NL}}}} \right)}^2}}{{Mt{\rho _{{{\rm{p}}_{nk}}}}}}\right).\nonumber
\end{align}

\section*{Appendix C\\\ Proof of Lemma 3}
Given the definition of ${Z_{nk}} = {{\left\| {({\boldsymbol{h}}_{nnkk}^{\rm{L}})^{\rm{H}} + {\boldsymbol{e}}_{nk}^{\rm{H}}} \right\|}^2}$, we define
\begin{equation}
{\boldsymbol{z}}_{nk}={({\boldsymbol{h}}_{nnkk}^{\rm{L}})^{\rm{H}} + {\boldsymbol{e}}_{nk}^{\rm{H}}}
= ({\boldsymbol{h}}_{nnkk}^{\rm{L}})^{\rm{H}} +  \sum\limits_{l \ne n}^{N} {\sqrt {\frac{{{\rho _{{{\rm{p}}_{lk}}}}}}{{{\rho _{{\rm{p}}_{nk}}}}}} \left( {{{{\boldsymbol{\bar h}}}_{lnkk}^{\rm{H}}} + {{{\boldsymbol{\tilde h}}}_{lnkk}^{\rm{H}}}} \right)}  + \frac{1}{{\sqrt {t{\rho _{{\rm{p}}_{nk}}}} }}{{\boldsymbol{w}}_{nk}^{\rm{H}}}, 
\end{equation}
where ${{\boldsymbol{z}}_{nk}} \in \mathbb{C}{^{M}} = \left[z_{nk1},\cdots,z_{nkM} \right]$ and
\begin{equation}
z_{nkm} = \beta _{nnkkm}^{\rm{L}}{h_{nnkkm}^*} + 
\sum\limits_{l \ne n}^{N} \sqrt {\frac{{{\rho _{{{\rm{p}}_{lk}}}}}{{{{\kappa _{lnkk}}}}}}{{{\rho _{{\rm{p}}_{nk}}}\left({{{\kappa _{lnkk}} + 1}}\right)}}}\left({ \beta _{lnkkm}^{\rm{L}}{h_{lnkkm}^{*}}}+ 
 \frac{{{\boldsymbol{g}}_{lnkk}^{\rm{H}}}{{\boldsymbol{r}}_{lnkkm}^{\rm{H}}}}{\sqrt {{{{\kappa _{lnkk}}}}}}\right)
+\frac{1}{{\sqrt {t{\rho _{{\rm{p}}_{nk}}}} }}{{{w}}_{nkm}^{*}}. \nonumber
\end{equation}
Since ${{{\boldsymbol{g}}_{lnkk}^{\rm{H}}}{{\boldsymbol{r}}_{lnkkm}^{\rm{H}}}}$ is calculated by the sum of $P$ independent complex Gaussian random variables,
${{{\boldsymbol{g}}_{lnkk}^{\rm{H}}}{{\boldsymbol{r}}_{lnkkm}^{\rm{H}}}}$ is also a complex Gaussian random variable.
Thus, we have $z_{nkm}\sim \mathcal{CN}\left(\mu _{{z_{nkm}}}, \sigma _{{z_{nkm}}}^2\right)$, where
\begin{align}
\mu _{{z_{nkm}}} &= \beta _{nnkkm}^{\rm{L}}{h_{nnkkm}^*} + \sum\limits_{l \ne n}^{N} {\sqrt {\frac{{{\rho _{{{\rm{p}}_{lk}}}}{\kappa _{lnkk}}}}{{\rho _{{{\rm{p}}_{nk}}}}\left({{\kappa _{lnkk}} + 1}\right)}} \beta _{lnkkm}^{\rm{L}}{h_{lnkkm}^*}},\label{eq.C.Msznkm}\\
\sigma _{{z_{nkm}}}^2  &= \sum\limits_{l \ne n}^{N} {\frac{{{\rho_{{\rm{p}}_{lk}}}{\left\| {\boldsymbol{r}}_{lnkkm} \right\|}^2 }}{{{\rho _{{{\rm{p}}_{nk}}}}\left({{{\kappa _{lnkk}} + 1}}\right)}}}  + \frac{1}{{t{\rho _{{{\rm{p}}_{nk}}}}}}. \label{eq.C.Sznkm}
\end{align}
From the definition of ${Z_{nk}} = {{\left\| {\boldsymbol{z}}_{nk} \right\|}^2}$, we finally have
${{\mu} _{{Z_{nk}}}}=\sum\limits_{m=1}^M\left({\sigma _{{z_{nkm}}}^2 + \left| {{\mu _{{z_{nkm}}}}}\right|^2}\right).$

\vspace{-0.3cm}
\section*{Appendix D\\ Proof of Lemma 4}\vspace{-0.15cm}
Given the definition of ${I_{nk}} $ from (\ref{eq.Ink}), we have
\begin{align}
\frac{I_{nk}}{M^2} &= {{{{\rho _{nk}}{{\left| \frac{{\boldsymbol{e}}_{nk}^{\rm{H}}{{\boldsymbol{h}}_{nnkk}^{\rm{L}}}}{M} \right|}^2} + \sum\limits_{j \ne k}^K {{\rho _{nj}}{{\left| \frac{{\boldsymbol{\hat h}}_{nnkk}^{\rm{H}}{{\boldsymbol{h}}_{nnjk}}}{M} \right|}^2}}  + \sum\limits_{l \ne n}^N {\sum\limits_{j = 1}^K {{\rho _{lj}}{{\left| \frac{{\boldsymbol{\hat h}}_{nnkk}^{\rm{H}}{{\boldsymbol{h}}_{lnjk}}}{M} \right|}^2} + {{\left\| \frac{({\boldsymbol{h}}_{nnkk}^{\rm{L}})^{\rm{H}} + {\boldsymbol{e}}_{nk}^{\rm{H}}}{M} \right\|}^2}} } }}} \nonumber\\
&= {{{{\rho _{nk}}{{\left| \frac{x_{nk}}{M} \right|}^2} + \sum\nolimits_{j \ne k}^K {{\rho _{nj}}{{\left| \frac{y_{njk}}{M} \right|}^2}}  + \sum\nolimits_{l \ne n}^N {\sum\nolimits_{j = 1}^K {{\rho _{lj}}{{\left| \frac{y_{lnjk}}{M} \right|}^2} + {{\left\| \frac{{\boldsymbol{z}}_{nk}}{M} \right\|}^2}} } }}}. 
\end{align}
In order to analyze the scaling law of ${\sigma _{{I_{nk}}}^2}/M^4$, we determine the scaling laws of $\sigma_{x_{nk}}^2/{M^2}$, $\sigma_{y_{njk}}^2/{M^2}$, $\sigma_{y_{lnjk}}^2/{M^2}$, and $\sigma_{z_{nkm}}^2/{M^2}$ according to $M$.
First, we determine the scaling law of $\sigma_{x_{nk}}^2/{M^2}$ from (\ref{eq.A.Sxnk}).
From (\ref{eq.DNLOS}), the correlation vector ${\boldsymbol{c}}_{lnkkp}$ is normalized by $\sqrt{M}$
and ${\boldsymbol{c}}_{lnkkp}^{\rm{H}}{\boldsymbol{h}}_{nnkk}^{\rm{L}}$ in (\ref{eq.A.Sxnk}) is calculated by the sum of $M$ elements.
Then, both ${\left| {{\boldsymbol{c}}_{lnkkp}^{\rm{H}}{{\boldsymbol{h}}_{nnkk}^{\rm{L}}}} \right|}^2$ and $\sum\nolimits_{m = 1}^M {\beta _{nnkkm}^2}$ in (\ref{eq.A.Sxnk}) increase with $\mathcal{O}({M})$ 
and thus, $\sigma_{x_{nk}}^2/{M^2}$ decreases with $\mathcal{O}(1/{M})$ as $M$ increases.
Therefore, we have $\sigma_{x_{nk}}^2/{M^2} \xrightarrow[\hspace{-2pt}M \to \infty \hspace{-2pt}]{}0$.
Next, we analyze $\sigma_{y_{njk}}^2/{M^2}$ from (\ref{eq.B.Synjk}) for large $M$.
From (\ref{eq.B.ynjk_EL}), $\sigma _{{y_{njk}^{\rm{EL}}}}^2$ follows $\mathcal{O}({M})$ similarly as $\sigma_{x_{nk}}^2$, and therefore, $\sigma _{{y_{njk}^{\rm{EL}}}}^2/M^2$ follows $\mathcal{O}({1/M})$ as $M$ increases.
To determine the scaling law of $\sigma _{{y_{njk}^{\rm{EN}}}}^2/M^2$,
we analyze the terms ${\left| {{{{\boldsymbol{\bar q}}}_{nk}^{\rm{H}}}{\boldsymbol{c}}_{nnjkp}} \right|}^2$, ${\big\| {{{\big( {{\boldsymbol{R}}_{lnkk}^{1/2}} \big)}^{\rm{H}}}{\boldsymbol{R}}_{nnjk}^{1/2}} \big\|_{\rm{F}}^2}$, and $\sum\nolimits_{m,p}^{M,P} {{{\left( {\alpha _{nnjkp}^{{\rm{NL}}}l_{nnjkm}^{{\rm{NL}}}} \right)}^2}}/M$ in (\ref{eq.B.ynjk_EN}) for large $M$.
Given that the correlation vector, ${\boldsymbol{c}}_{nnjkp}$, and matrices, ${{\boldsymbol{R}}_{lnkk}^{1/2}}$ and ${{\boldsymbol{R}}_{nnjk}^{1/2}}$, are normalized by $\sqrt{M}$,
the terms ${\left| {{{{\boldsymbol{\bar q}}}_{nk}^{\rm{H}}}{\boldsymbol{c}}_{nnjkp}} \right|}^2$, ${\big\| {{{\big( {{\boldsymbol{R}}_{lnkk}^{1/2}} \big)}^{\rm{H}}}{\boldsymbol{R}}_{nnjk}^{1/2}} \big\|_{\rm{F}}^2}$, and $\sum\nolimits_{m,p}^{M,P} {{{\left( {\alpha _{nnjkp}^{{\rm{NL}}}l_{nnjkm}^{{\rm{NL}}}} \right)}^2}}/M$
follow, respectively, $\mathcal{O}({M})$, $\mathcal{O}({1})$, and $\mathcal{O}({1})$ as $M$ increases.
Consequently, $\sigma _{{y_{njk}^{\rm{EN}}}}^2/M^2$ decreases with $\mathcal{O}({1/M})$ and we have $\sigma_{y_{njk}}^2/{M^2} \xrightarrow[\hspace{-2pt}M \to \infty \hspace{-2pt}]{}0$.
Similarly, $\sigma_{y_{lnjk}}^2/{M^2}$ from (\ref{eq.B.Sylnjk}) also converges to zero as $M \to \infty$.
Finally, we determine the scaling law of $\sigma_{z_{nkm}}^2/{M^2}$ from (\ref{eq.C.Sznkm}).
Given that ${\boldsymbol{r}}_{lnkkm}$ is a correlation vector normalized by $\sqrt{M}$,
${\left\| {\boldsymbol{r}}_{lnkkm} \right\|}^2$ is calculated by the sum of $P$ elements divided by $M$.
Hence, $\sigma_{z_{nkm}}^2/{M^2}$ decreases with $\mathcal{O}({1/M^3})$ as $M$ increases,
and the variance of ${{\left\| {{\boldsymbol{z}}_{nk}}/{M} \right\|}^2}$ eventually converges to zero as $M \to \infty$.
In conclusion, we have ${\sigma _{{I_{nk}}}^2}/M^4\xrightarrow[\hspace{-2pt}M \to \infty \hspace{-2pt}]{}0$, which completes the proof.

\vspace{-0.2cm}
\section*{Appendix E\\ Proof of Theorem 1}\vspace{-0.15cm}
We begin with the definition of $R_{nk}$ as follows: 
\begin{equation}
{{R}}_{nk} 
= {\log }\left( {1 + \frac{{{\rho _{nk}}{S_{nk}}}}{{{I_{nk}}}}}  \right)  ={{{\log }}\left( {{\rho _{nk}}{{S_{nk}}}}+{{I_{nk}}} \right)} - {{{\log\rm{ } }} {{I_{nk}}}}.
\end{equation}
For notational simplicity, we define ${{R_{nk}^{\rm{L}}}}={{{\log }}\left( {{\rho _{nk}}{{S_{nk}}}}+{{I_{nk}}} \right)}$ and ${{R_{nk}^{\rm{R}}}}={{\log\rm{ } }} {{I_{nk}}}$.
Then, we have 
\begin{equation}
{{R_{nk}^{\rm{L}}}} 
= \frac{ {{{I_{nk}}} - {\bar \mu _{{{I_{nk}}}}} } }{M^2}{\log }{\left( {1 + \frac{\left({{I_{nk}} - {\bar \mu _{{I_{nk}}}} }\right)/M^2}{\left({{{\rho _{nk}}{{S_{nk}}}}  + {\bar \mu _{{{I_{nk}}}}} }\right)/M^2}}\right)^{\frac{M^2}{{{{I_{nk}}} - {\bar \mu _{{{I_{nk}}}}} }}}}
+ {\log }\left( {{{\rho _{nk}}{S_{nk}}}  + {\bar \mu _{{I_{nk}}}} } \right).
\end{equation}
Since ${{I_{nk}}}/M^2-{\bar \mu _{{I_{nk}}}}/M^2 \xrightarrow[M \to \infty ]{} 0$ from Lemma 4
and ${S_{nk}} - {\bar p_{nk}} \xrightarrow[M \to \infty ]{} 0$ where ${{\bar p}_{nk}} = {{{M^2}p_{nk}^2}}$,
we have the following asymptotic convergence using the exponential function definition $e^x = \mathop {\lim }\limits_{n \to \infty } {\left( {1 + x/n} \right)^n}$: 
${{R_{nk}^{\rm{L}}}} - {{{\bar R}_{nk}^{\rm{L}}}}\xrightarrow[M \to \infty ]{} 0$, where
\begin{equation}
{{{\bar R}_{nk}^{\rm{L}}}}\mathop  = 
 \left( {\frac{{{{I_{nk}}} - {\bar \mu _{{{I_{nk}}}}} }}{{{{\rho _{nk}}{{\bar p_{nk}}}}  + {\bar \mu _{{{I_{nk}}}}} }}} \right) + {\log }\left( {{{\rho _{nk}}{{\bar p_{nk}}}}  + {\bar \mu _{{{I_{nk}}}}} } \right). \label{eq.RA}
\end{equation}
Similarly, we have ${{R_{nk}^{\rm{R}}}} - {{{\bar R}_{nk}^{\rm{R}}}}\xrightarrow[\hspace{-2pt}M \to \infty \hspace{-2pt}]{} 0$, where
\begin{equation}
{{R_{nk}^{\rm{R}}}} 
= \frac{ {{I_{nk}}-  {\bar \mu _{{{I_{nk}}}}} } }{M^2}{\log }{\left( {1 + \frac{\left({{{I_{nk}}} -  {\bar \mu _{{{I_{nk}}}}} }\right)/M^2}{ {\bar \mu _{{{I_{nk}}}}}/M^2}} \right)^{\frac{M^2}{{{{I_{nk}}} -  {\bar \mu _{{{I_{nk}}}}} }}}} + {\log\rm{ }} {\bar \mu _{{{I_{nk}}}}},
\end{equation}
\begin{equation}
{{{\bar R}_{nk}^{\rm{R}}}} = {\frac{{{I_{nk}} -  {\bar \mu _{{I_{nk}}}}}}{ {\bar \mu _{{{I_{nk}}}}} }} + {\log\rm{ }} {\bar \mu _{{I_{nk}}}}.\label{eq.RB}
\end{equation}
From (\ref{eq.RA}) and (\ref{eq.RB}), we thus have
\begin{equation}
{{{\bar R}_{nk}^{\rm{L}}}} - {{{\bar R}_{nk}^{\rm{R}}}}
= \left( {1 - \frac{{{I_{nk}}}}{{{{\bar \mu }_{{I_{nk}}}}}}} \right)\frac{{{\rho _{nk}}{{\bar p_{nk}}}}}{{{\rho _{nk}}{{\bar p_{nk}}} + {{\bar \mu }_{{I_{nk}}}}}} + \log \left( {1 + \frac{{{\rho _{nk}}{{\bar p_{nk}}}}}{{{{\bar \mu }_{{I_{nk}}}}}}} \right). 
\end{equation}
Given that ${I_{nk}}/{M^2} -  {{\bar\mu} _{{I_{nk}}}}/{M^2}\xrightarrow[M \to \infty]{\rm{}}0$,
${\bar{R}}_{nk}$ can be derived as follows:
\begin{equation}
{\bar R}_{nk} =\log \left( {1 + \frac{{{\rho _{nk}}{{\bar p_{nk}}}}}{{{{\bar \mu }_{{I_{nk}}}}}}} \right)= {\log \left( {1 + \frac{{M^2{\rho _{nk}}p_{nk}^2}}{{{\bar \mu}_{I_{nk}}}}} \right)},
\end{equation}
where ${{R_{nk}}} - {{{\bar R}_{nk}}}\xrightarrow[\hspace{-2pt}M \to \infty \hspace{-2pt}]{} 0$,
which completes the proof.

\bibliographystyle{IEEEtran}
\bibliography{IEEEabrv,myBiB}

\begin{thebibliography}{10}
\providecommand{\url}[1]{#1}
\csname url@samestyle\endcsname
\providecommand{\newblock}{\relax}
\providecommand{\bibinfo}[2]{#2}
\providecommand{\BIBentrySTDinterwordspacing}{\spaceskip=0pt\relax}
\providecommand{\BIBentryALTinterwordstretchfactor}{4}
\providecommand{\BIBentryALTinterwordspacing}{\spaceskip=\fontdimen2\font plus
\BIBentryALTinterwordstretchfactor\fontdimen3\font minus
  \fontdimen4\font\relax}
\providecommand{\BIBforeignlanguage}[2]{{%
\expandafter\ifx\csname l@#1\endcsname\relax
\typeout{** WARNING: IEEEtran.bst: No hyphenation pattern has been}%
\typeout{** loaded for the language `#1'. Using the pattern for}%
\typeout{** the default language instead.}%
\else
\language=\csname l@#1\endcsname
\fi
#2}}
\providecommand{\BIBdecl}{\relax}
\BIBdecl

\bibitem{ref.Jung2019PCcon}
M.~{Jung}, W.~{Saad}, and G.~{Kong}, ``Uplink spectral efficiency in large
  intelligent surfaces: {A}symptotic analysis under pilot contamination,'' in
  \emph{IEEE GLOBECOM}, Waikoloa, USA, Dec. 2019.

\bibitem{ref.Basar2019indexmodulation}
E.~Basar, ``Large intelligent surface-based index modulation: A new beyond
  {MIMO} paradigm for 6{G},'' \emph{available online:
  arxiv.org/abs/1904.06704}, Apr. 2019.

\bibitem{ref.Saad20196G}
W.~{Saad}, M.~{Bennis}, and M.~{Chen}, ``A vision of 6{G} wireless systems:
  {A}pplications, trends, technologies, and open research problems,''
  \emph{IEEE Network, to appear}, 2019.

\bibitem{ref.Ericsson2011IoT}
{Ericsson White Paper}, ``More than 50 billion connected devices,'' Ericsson,
  Stockholm, Sweden, Tech. Rep. 284 23-3149 Uen, Feb. 2011.

\bibitem{ref.Dawy2017MTC}
Z.~{Dawy}, W.~{Saad}, A.~{Ghosh}, J.~{Andrews}, and E.~{Yaacoub}, ``Towards
  massive machine type cellular communications,'' \emph{{IEEE} Commun. Mag.},
  vol.~24, no.~1, pp. 120--128, Feb. 2017.

\bibitem{ref.Tae2016learning}
T.~{Park}, N.~{Abuzainab}, and W.~{Saad}, ``Learning how to communicate in the
  {I}nternet of {T}hings: Finite resources and heterogeneity,'' \emph{{IEEE}
  Access}, vol.~4, pp. 7063--7073, Nov. 2016.

\bibitem{ref.Mozaffari2019beyond}
M.~{Mozaffari}, A.~T.~Z. {Kasgari}, W.~{Saad}, M.~{Bennis}, and M.~{Debbah},
  ``Beyond {5G} with {UAV}s: {F}oundations of a {3D} wireless cellular
  network,'' \emph{{IEEE} Trans. Wireless Commun.}, vol.~18, no.~1, pp.
  357--372, Jan. 2019.

\bibitem{ref.Mozaffari2017mobile}
M.~{Mozaffari}, W.~{Saad}, M.~{Bennis}, and M.~{Debbah}, ``Mobile unmanned
  aerial vehicles ({UAV}s) for energy-efficient internet of things
  communications,'' \emph{{IEEE} Trans. Wireless Commun.}, vol.~16, no.~11, pp.
  7574--7589, Nov. 2017.

\bibitem{ref.Zeng2018joint}
T.~{Zeng}, O.~{Semiari}, W.~{Saad}, and M.~{Bennis}, ``Joint communication and
  control for wireless autonomous vehicular platoon systems,'' \emph{available
  online: arxiv.org/abs/1804.05290}, Apr. 2018.

\bibitem{ref.Jung2018lisul}
M.~{Jung}, W.~{Saad}, Y.~{Jang}, G.~{Kong}, and S.~{Choi}, ``Performance
  analysis of large intelligent surfaces ({LIS}s): {A}symptotic data rate and
  channel hardening effects,'' \emph{{IEEE} Trans. Wireless Commun.}, vol.~19,
  no.~3, pp. 2052--2065, Mar. 2020.

\bibitem{ref.Hu2018data}
S.~{Hu}, F.~{Rusek}, and O.~{Edfors}, ``Beyond massive {MIMO}: The potential of
  data transmission with large intelligent surfaces,'' \emph{{IEEE} Trans.
  Signal Process.}, vol.~66, no.~10, pp. 2746--2758, May 2018.

\bibitem{ref.Hu2018assignment}
S.~{Hu}, K.~{Chitti}, F.~{Rusek}, and O.~{Edfors}, ``User assignment with
  distributed large intelligent surface ({LIS}) systems,'' in \emph{IEEE
  PIMRC}, Bologna, Italy, Sep. 2018.

\bibitem{ref.Jung2019reliability}
M.~Jung, W.~Saad, Y.~Jang, G.~Kong, and S.~Choi, ``Reliability analysis of
  large intelligent surfaces ({LIS}s): Rate distribution and outage
  probability,'' \emph{IEEE Wireless Communications Letters}, vol.~8, no.~6,
  pp. 1662--1666, Dec. 2019.

\bibitem{ref.Hu2018positioning}
S.~{Hu}, F.~{Rusek}, and O.~{Edfors}, ``Beyond massive {MIMO}: The potential of
  positioning with large intelligent surfaces,'' \emph{{IEEE} Trans. Signal
  Process.}, vol.~66, no.~7, pp. 1761--1774, Apr. 2018.

\bibitem{ref.Hu2018hardware}
------, ``Capacity degradation with modeling hardware impairment in large
  intelligent surface,'' \emph{available online: arxiv.org/abs/1810.09672},
  Oct. 2018.

\bibitem{ref.LISnew}
J.~Yuan, H.~Q. Ngo, and M.~Matthaiou, ``Large intelligent surface ({LIS})-based
  communications: New features and system layouts,'' \emph{available online:
  arxiv.org/abs/2002.12183}, Feb. 2020.

\bibitem{ref.Han2018assisted}
Y.~{Han}, W.~{Tang}, S.~{Jin}, C.~{Wen}, and X.~{Ma}, ``Large intelligent
  surface-assisted wireless communication exploiting statistical {CSI},''
  \emph{available online: arxiv.org/abs/1812.05429}, Dec. 2018.

\bibitem{ref.Huang2018energy}
C.~{Huang}, A.~{Zappone}, G.~C. {Alexandropoulos}, M.~{Debbah}, and C.~{Yuen},
  ``Reconfigurable intelligent surfaces for energy efficiency in wireless
  communication,'' \emph{{IEEE} Trans. Wireless Commun.}, vol.~18, no.~8, pp.
  4157--4170, Aug 2019.

\bibitem{ref.Wu2018beamforming}
Q.~{Wu} and R.~{Zhang}, ``Intelligent reflecting surface enhanced wireless
  network via joint active and passive beamforming,'' \emph{IEEE Transactions
  on Wireless Communications}, vol.~18, no.~11, pp. 5394--5409, Nov. 2019.

\bibitem{ref.Wu2019bfoptimization}
------, ``Beamforming optimization for intelligent reflecting surface with
  discrete phase shifts,'' in \emph{IEEE ICASSP}, Brighton, UK, May 2019.

\bibitem{ref.editor}
L.~{Dai}, B.~{Wang}, M.~{Wang}, X.~{Yang}, J.~{Tan}, S.~{Bi}, S.~{Xu},
  F.~{Yang}, Z.~{Chen}, M.~D. {Renzo}, C.~{Chae}, and L.~{Hanzo},
  ``Reconfigurable intelligent surface-based wireless communications: {A}ntenna
  design, prototyping, and experimental results,'' \emph{IEEE Access}, vol.~8,
  pp. 45\,913--45\,923, 2020.

\bibitem{ref.PSS00}
M.~Cui, G.~Zhang, and R.~Zhang, ``Secure wireless communication via intelligent
  reflecting surface,'' \emph{IEEE Wireless Communications Letters}, vol.~8,
  no.~5, pp. 1410--1414, Oct. 2019.

\bibitem{ref.WPT00}
C.~Pan, H.~Ren, K.~Wang, M.~Elkashlan, A.~Nallanathan, J.~Wang, and L.~Hanzo,
  ``Intelligent reflecting surface enhanced {MIMO} broadcasting for
  simultaneous wireless information and power transfer,'' \emph{available
  online: arxiv.org/abs/1908.04863}, Feb. 2020.

\bibitem{ref.WPT01}
Q.~Wu and R.~Zhang, ``Joint active and passive beamforming optimization for
  intelligent reflecting surface assisted {SWIPT} under {QoS} constraints,''
  \emph{IEEE Journal of Selected Areas in Communications}, {to appear,} 2020.

\bibitem{ref.RR1}
C.~{Huang}, S.~{Hu}, G.~C. {Alexandropoulos}, A.~{Zappone}, C.~{Yuen},
  R.~{Zhang}, M.~D. {Renzo}, and M.~{Debbah}, ``Holographic {MIMO} surfaces for
  6{G} wireless networks: {O}pportunities, challenges, and trends,'' \emph{IEEE
  Wireless Communications}, vol.~27, no.~5, pp. 118--125, 2020.

\bibitem{ref.RR2}
L.~Wei, C.~Huang, G.~C. Alexandropoulos, C.~Yuen, Z.~Zhang, and M.~Debbah,
  ``Channel estimation for {RIS}-empowered multi-user {MISO} wireless
  communications,'' \emph{available online: arxiv.org/abs/2008.01459}, Aug.
  2020.

\bibitem{ref.RR3}
C.~{Huang}, R.~{Mo}, and C.~{Yuen}, ``Reconfigurable intelligent surface
  assisted multiuser {MISO} systems exploiting deep reinforcement learning,''
  \emph{IEEE Journal on Selected Areas in Communications}, vol.~38, no.~8, pp.
  1839--1850, 2020.

\bibitem{ref.RR4}
J.~{Gao}, C.~{Zhong}, X.~{Chen}, H.~{Lin}, and Z.~{Zhang}, ``Unsupervised
  learning for passive beamforming,'' \emph{IEEE Communications Letters},
  vol.~24, no.~5, pp. 1052--1056, 2020.

\bibitem{ref.DMA1}
N.~Shlezinger, G.~C. Alexandropoulos, M.~F. Imani, Y.~C. Eldar, and D.~R.
  Smith, ``Dynamic metasurface antennas for 6{G} extreme massive {MIMO}
  communications,'' \emph{available online: arxiv.org/abs/2006.07838}, Jun.
  2020.

\bibitem{ref.DMA2}
N.~Shlezinger, O.~Dicker, Y.~C. Eldar, I.~Yoo, M.~F. Imani, and D.~R. Smith,
  ``Dynamic metasurface antennas for uplink massive {MIMO} systems,''
  \emph{available online: arxiv.org/abs/1901.01458}, Jun. 2019.

\bibitem{ref.Kim2018scaling}
T.~{Kim}, K.~{Min}, M.~{Jung}, and S.~{Choi}, ``Scaling laws of optimal
  training lengths for {TDD} massive {MIMO} systems,'' \emph{{IEEE} Trans. Veh.
  Technol.}, vol.~67, no.~8, pp. 7128--7142, Aug. 2018.

\bibitem{ref.Hoydis2013how}
J.~{Hoydis}, S.~{ten Brink}, and M.~{Debbah}, ``Massive {MIMO} in the {UL/DL}
  of cellular networks: How many antennas do we need?'' \emph{{IEEE} J. Sel.
  Areas Commun.}, vol.~31, no.~2, pp. 160--171, Feb. 2013.

\bibitem{ref.Marzetta2010noncooperative}
T.~L. {Marzetta}, ``Noncooperative cellular wireless with unlimited numbers of
  base station antennas,'' \emph{{IEEE} Trans. Wireless Commun.}, vol.~9,
  no.~11, pp. 3590--3600, Nov. 2010.

\bibitem{ref.scaling1}
L.~Yang, Y.~Yang, D.~B. da~Costa, and I.~Trigui, ``Outage probability and
  capacity scaling law of multiple ris-aided cooperative networks,''
  \emph{available online: arxiv.org/abs/2007.13293}, Jul. 2020.

\bibitem{ref.scaling2}
K.~Zhi, C.~Pan, H.~Ren, and K.~Wang, ``Power scaling law analysis and phase
  shift optimization of ris-aided massive mimo systems with statistical csi,''
  \emph{available online: arxiv.org/abs/2010.13525}, Dec. 2020.

\bibitem{ref.LTE2017TR36211}
3rd Generation Partnership~Project, ``{Technical Specification Group Radio
  Access Network; Physical channels and modulation},'' TR 36.211, {V15.0.0},
  Dec. 2017.

\bibitem{ref.Miao2016mobile}
G.~{Miao}, J.~{Zander}, K.~{Sung}, and B.~{Slimane}, \emph{Fundamentals of
  Mobile Data Networks}.\hskip 1em plus 0.5em minus 0.4em\relax Cambridge Univ.
  Press, 2016.

\bibitem{ref.Tse2005fundamentals}
D.~Tse and P.~Viswanath, \emph{Fundamentals of Wireless Communication}.\hskip
  1em plus 0.5em minus 0.4em\relax Cambridge Univ. Press, 2005.

\bibitem{ref.Song2017UPA}
J.~{Song}, J.~{Choi}, and D.~J. {Love}, ``Common codebook millimeter wave beam
  design: Designing beams for both sounding and communication with uniform
  planar arrays,'' \emph{{IEEE} Trans. Commun.}, vol.~65, no.~4, pp.
  1859--1872, Apr. 2017.

\bibitem{ref.Han2014design}
Y.~{Han}, S.~{Jin}, X.~{Li}, Y.~{Huang}, L.~{Jiang}, and G.~{Wang}, ``Design of
  double codebook based on 3{D} dual-polarized channel for multiuser {MIMO}
  system,'' \emph{EURASIP J. Adv. Signal Process.}, vol. 2014, no.~1, pp.
  1--13, Jul. 2014.

\bibitem{ref.Jose2011PC}
J.~{Jose}, A.~{Ashikhmin}, T.~L. {Marzetta}, and S.~{Vishwanath}, ``Pilot
  contamination and precoding in multi-cell {TDD} systems,'' \emph{{IEEE}
  Trans. Wireless Commun.}, vol.~10, no.~8, pp. 2640--2651, Aug. 2011.

\bibitem{ref.Khansefid2015LS}
A.~{Khansefid} and H.~{Minn}, ``On channel estimation for massive {MIMO} with
  pilot contamination,'' \emph{{IEEE} Commun. Lett.}, vol.~19, no.~9, pp.
  1660--1663, Sep. 2015.

\bibitem{ref.R2A4}
Q.~{Wu} and R.~{Zhang}, ``Towards smart and reconfigurable environment:
  Intelligent reflecting surface aided wireless network,'' \emph{IEEE
  Communications Magazine}, vol.~58, no.~1, pp. 106--112, 2020.

\bibitem{ref.Poor1994introduction}
H.~V. Poor, \emph{An Introduction to Signal Detection and Estimation}.\hskip
  1em plus 0.5em minus 0.4em\relax Springer, 1994.

\bibitem{ref.Boyd2004convex}
S.~{Boyd} and L.~{Vandenberghe}, \emph{Convex Optimization}.\hskip 1em plus
  0.5em minus 0.4em\relax Cambridge Univ. Press, 2004.

\bibitem{ref.LTE2017TR25996}
3rd Generation Partnership~Project, ``{Technical Specification Group Radio
  Access Network; Spatial channel model for Multiple Input Multiple Output
  (MIMO) simulations},'' TR 25.996, {V14.0.0}, Mar. 2017.

\bibitem{ref.LTE2018TR36331}
------, ``{Technical Specification Group Radio Access Network; Radio Resource
  Control (RRC); Protocol specification},'' TR 36.331, {V15.0.1}, Jan. 2018.

\end{thebibliography}

\end{document}